\pdfoutput=1
\documentclass{aa}
\usepackage{txfonts}
\usepackage{natbib}
\usepackage{graphicx}
\bibpunct{(}{)}{;}{a}{}{,}
\usepackage{color}

\begin{document}

\title{Infrared photometry and CaT spectroscopy of globular cluster M\,28 (NGC\,6626)\thanks{Based on observations gathered
at the European Southern Observatory (programme ID 091.D-0535A), and with ESO-VISTA telescope (programme ID 172.B-2002).}
}
\subtitle{}

\author{
C. Moni Bidin\inst{1}
\and
F. Mauro\inst{1}
\and
R. Contreras Ramos\inst{2,3}
\and
M. Zoccali\inst{2,3}
\and
Y. Reinarz\inst{1}
\and
M. Moyano\inst{1}
\and
D. Gonz\'alez-D\'iaz\inst{1,4}
\and
S. Villanova\inst{5}
\and
G. Carraro\inst{6}
\and
J. Borissova\inst{7,2}
\and
A.-N. Chen\'e\inst{8}
\and
R. E. Cohen\inst{9}
\and
D. Geisler\inst{5,10,11}
\and
R. Kurtev\inst{7,2}
\and
D. Minniti\inst{12,13}
}

\institute{
Instituto de Astronom\'ia, Universidad Cat\'olica del Norte, Av. Angamos 0610, Antofagasta, Chile
\and
Millennium Institute of Astrophysics, Av. Vicu\~na Mackenna 4860, 782-0436 Macul, Santiago, Chile
\and
Instituto de Astrof\'isica, Pontificia Universidad Cat\'olica de Chile, Av. Vicu\~na Mackenna 4860,
782-0436 Macul, Santiago, Chile
\and
Instituto de F\'isica-FCEN, Universidad de Antioquia, Calle 70 No. 52-21, Medellin, Colombia
\and
Departamento de Astronom\'ia, Universidad de Concepci\'on, Casilla 160-C, Concepci\'on, Chile
\and
Dipartimento di Fisica e Astronomia, Universit\'a di Padova, Vicolo Osservatorio 3 I-35122, Padova, Italy
\and
Instituto de F\'isica y Astronom\'ia, Universidad de Valpara\'iso, Av. Gran Breta\~na 1111, Playa Ancha, Casilla 5030, Chile.
\and
Gemini Observatory/NSF’s NOIRLab, 670 N. A`ohoku Place, Hilo, HI 96720, USA
\and
Space Telescope Science Institute, 3700 San Martin Drive, Baltimore, MD 21218, USA
\and
Instituto de Investigaci\'on Multidisciplinario en Ciencia y Tecnolog\'ia, Universidad de La Serena. Avenida Ra\'ul Bitr\'an
S/N, La Serena, Chile
\and
Departamento de F\'isica y Astronom\'ia, Facultad de Ciencias, Universidad de La Serena. Av. Juan Cisternas 1200, La Serena, Chile
\and
Departamento de F\'isica, Facultad de Ciencias Exactas, Universidad Andres Bello, Av. Fernandez Concha 700,
Las Condes, Santiago, Chile
\and
Vatican Observatory, V00120 Vatican City State, Italy
}
\date{Received / Accepted }


\abstract
{Recent studies show that the inner Galactic regions host genuine bulge globular clusters, but also halo intruders, complex
remnants of primordial building blocks, and objects likely accreted during major merging events.}
{In this study we focus on the properties of M\,28, a very old and massive cluster currently located in the Galactic bulge.}
{We analysed wide-field infrared photometry collected by the VVV survey, VVV proper motions, and intermediate-resolution spectra
in the calcium triplet range for 113 targets in the cluster area.}
{Our results in general confirm previous estimates of the cluster properties available in the literature. We find no evidence of
differences in metallicity between cluster stars, setting an upper limit of $\Delta [\frac{\mathrm{Fe}}{\mathrm{H}}]<0.08$~dex to
any internal inhomogeneity. We confirm that M\,28 is one of the oldest objects in the Galactic bulge (13-14~Gyr). From this
result and the literature data, we find evidence of a weak age--metallicity relation among bulge globular clusters that suggests
formation and chemical enrichment. In addition, wide-field density maps show that M\,28 is tidally stressed and that it is losing
mass into the general bulge field.}
{Our study indicates that M\,28 is a genuine bulge globular cluster, but its very old age and its mass loss suggest that this
cluster could be the remnant of a larger structure, possibly a primeval bulge building block.}


\authorrunning{Moni Bidin et al.}
\mail{cmoni@ucn.cl}
\titlerunning{Infrared photometry and CaT spectroscopy of M\,28}
\maketitle


\section{Introduction}
\label{s_intro}

Our understanding of the complexity of Galactic globular clusters (GCs) has expanded impressively in the last decade after the
discovery that they can host multiple stellar populations with different chemical enrichment histories
\citep[e.g.][]{Piotto15}.
The classical definition that they are simple stellar populations (i.e. initially chemically homogeneous aggregates of coeval
stars) is now outdated. \citet{Carretta09} showed that a certain degree of inhomogeneity of light chemical elements (C, N, O, Na,
Mg, and Al) is observed in all GCs, and the only exception confirmed so far is Ruprecht\,106 \citep{Villanova13}, and possibly
HP\,1 \citep{Barbuy16}. The coexistence of stars with different chemical compositions in the same cluster is believed to be due
to the very early evolution of the system, where a second generation \citep[or possibly more; e.g.][]{Villanova17} of stars formed
from gas polluted by matter chemically processed by the first cluster stars. Intermediate-mass asymptotic giant branch stars
\citep{DAntona02}, fast-rotating massive stars \citep{Decressin07}, and massive binaries \citep{deMink10} are the most accredited
candidate polluters, but none of these self-enrichment scenarios is free of problems \citep[see][]{Renzini15}, and their origin
is still hotly debated.

A spread in iron content is a characteristic restricted to only a few very massive globulars, such as $\omega$~Centauri
\citep[e.g.][]{Piotto05,Johnson08} and M\,54 \citep{Carretta10}. In the self-enrichment model this is interpreted as the
consequence of high initial cluster mass, enough to retain the highly accelerated ejecta of supernovae explosions
\citep[e.g.][]{Lee09}. \citet{Ferraro09} discovered the existence of two horizontal branches (HBs) in the bulge GC Terzan\,5, a
signature of the presence of two distinct stellar populations, with different spatial distributions, metallicities
\citep{Origlia11}, and possibly helium content \citep{DAntona10} and/or ages \citep{Ferraro09}. These peculiarities suggested that
we might be witnessing the relic of a large building block of the Galactic inner spheroid \citep{Lanzoni10}. Infrared data
collected by the Vista Variables in the V\'ia L\'actea (VVV) ESO public survey \citep{Minniti10} suggested a split HB even in
NGC\,6440 and NGC\,6569 \citep{Mauro12}. These two metal-rich GCs, as in Terzan\,5, are among the 10 most massive of the 64 GCs
within 4~kpc of the Galactic centre. More recent studies have excluded the presence of a significant metallicity spread in these
clusters \citep{Munoz17,Johnson18}. A spread in the content of heavy elements, however, is most likely present in M\,22, another
very massive GC lying in the inner regions of the Galaxy \citep{DaCosta09,Marino09,Marino11}, although this too is still debated
\citep{Mucciarelli15,Lee16}.

\citet{Minniti95b} first suggested that there is a population of GCs that are associated with the Milky Way bulge, on the basis of
metallicities, kinematics, and spatial distribution. M\,28 (NGC\,6626) is a poorly studied bulge GC located in the inner 3~kpc
of the Galactic centre. In many aspects it is similar to the extensively analysed M\,22, as it is also metal poor, is as massive as
M\,22 and NGC\,6569 \citep{Harris96,Peterson87}, and shows a blue and extended HB. Still, it must be noted that it is more
metal rich than M\,22 by 0.4~dex \citep[][hereafter V17]{Villanova17}, \defcitealias{Villanova17}{V17} and \citet{Bica16}
classified it as a genuine bulge cluster, while they considered M\,22 a likely Halo intruder. Few studies so far have estimated the
age of M\,28, but they all suggest that it should be a very old object \citep{Testa01,Roediger14,Villanova17,Kerber18}; indeed,
it might be one of the oldest objects in the Galactic bulge.

The aforementioned peculiarities, and its similarity with M\,22, make M\,28 a good candidate cluster to search for an internal
spread of iron. \citet{Prieto12} analysed the properties of RR~Lyrae variables in this cluster, and suggested the presence of
multiple stellar populations to explain their peculiar hybrid Oosterhoff behaviour. \citet[][hereafter M14]{Mauro14}
\defcitealias{Mauro14}{M14} re-analysed the calcium triplet (CaT) measurements of \citet[hereafter R97]{Rutledge97},
\defcitealias{Rutledge97}{R97} and the resulting metallicity distribution appears wide, with possibly two peaks separated by
$\sim$0.2~dex (see their Fig. 27). On the contrary, \citetalias{Villanova17} claimed no intrinsic spread of metallicity among
their 17 red giant branch (RGB) stars in M\,28 because the observed measured dispersion of $\sigma_\mathrm{[Fe/H]}$=0.06~dex equals
the estimated errors.

In this work we investigate M\,28 and its peculiarities further by means of VVV infrared photometry and low-resolution
spectroscopy. In particular, our aim is to estimate the cluster age, and to analyse the possibility of an internal spread of iron
via CaT spectroscopy of a large sample of stars.


\section{Observations and data reduction}
\label{s_data}

\subsection{Photometric data}
\label{ss_dataphot}

\begin{figure}
\begin{center}
\includegraphics[width=9.cm]{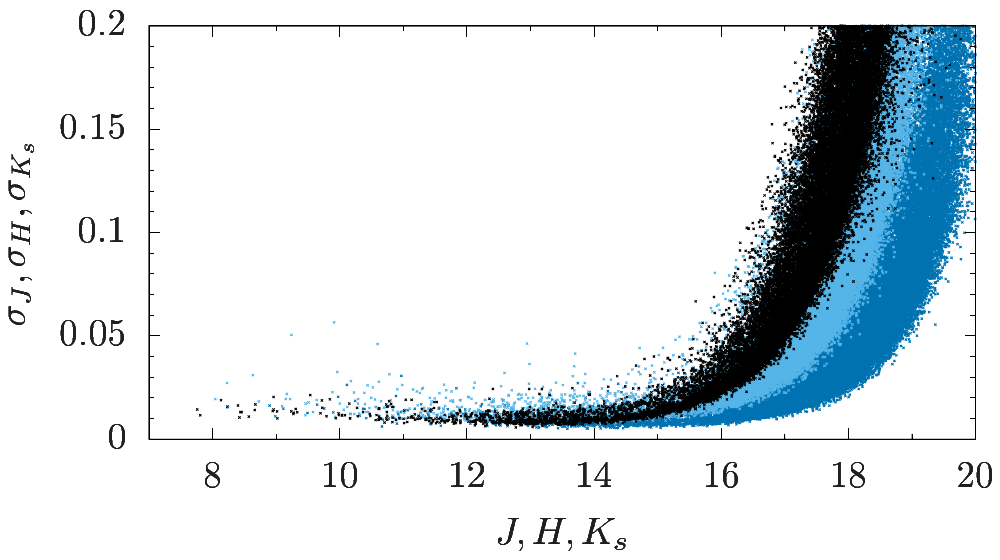}
\caption{Photometric errors as a function of magnitude in the $J$ (blue), $H$ (light blue), and $K_\mathrm{s}$ (black) bands.}
\label{f_photerr}
\end{center}
\end{figure}

The infrared (IR) photometry of M\,28 was performed on the data collected by the Vista Variables in the V\'ia L\'actea survey
\citep{Minniti10,Hempel14}. The data were acquired between 2010 July 1 and 2011 September 11 with the VIRCAM camera mounted on
the VISTA 4m telescope at the Paranal Observatory \citep{Emerson10}, and reduced at the Cambridge Astronomical Survey Unit
(CASU)\footnote{http://casu.ast.cam.ac.uk/} with the VIRCAM pipeline \citep{Irwin04}. During the observations the weather
conditions fell within the survey's constraints for seeing, airmass, and Moon distance \citep{Minniti10}, and the quality of
the data was satisfactory.

VIRCAM is an array of 16 independent 2048$\times$2048~pixel detectors. We retrieved from the Vista Science Archive
website\footnote{http://horus.roe.ac.uk/vsa/} the 175 frames collected by individual chips required to cover a wide field up
to $13\farcm5$ from the cluster centre, in the three $JHK_\mathrm{s}$ bands of the VVV survey. Each frame covered an area of
approximately $10\arcmin\times10\arcmin$, and overlapped with contiguous images so that any point in the field was sampled by
at least two frames. The retrieved data covered an area of $\sim40\arcmin\times30\arcmin$ in Galactic longitude and latitude,
respectively, with the cluster slightly off centre. The pixel scale was 0$\farcs$34, and the effective exposure time was 8s in
the $H$ and $K_\mathrm{s}$ bands, and 24s in $J$ band \citep[see][for a more detailed description of the VVV data]{Saito12}.

The PSF photometry was performed with the VVV-SkZ\_pipeline \citep{Mauro13}, based on the DAOPHOT~II and ALLFRAME codes
\citep{Stetson94}. The results were calibrated in the 2MASS astrometric and photometric system \citep{Skrutskie06}, as detailed
in \citet{Moni11} and \citet{Chene12}. Between 45 and 55 2MASS stars were used in each frame for this process, with magnitudes
in the range $H$=9.8--11.8, $J$=10.5--12.5, and $K_\mathrm{s}$=9.5--11.5. The VVV frames saturate at $K_\mathrm{s}\approx$12, but
\citet{Mauro13} showed that the PSF fit can recover correct magnitudes up to about two magnitudes above saturation. At brighter
magnitudes, the completeness progressively declines. Cluster stars brighter than $K_\mathrm{s}$=8.5 are nevertheless so scarce,
that we decided not to complement our VVV catalogue with the bright 2MASS sources present in the field. The trend of errors as
a function of magnitude is shown in Fig.~\ref{f_photerr}. The completeness of the VVV catalogue is limited by a lower detection
rate in the $J$ band, mainly because cluster stars are fainter in $J$ than in the other bands, due to their red colour. Defining
a successful detection only when the source is measured in all three bands, artificial star experiments on VVV frames showed
us that our PSF photometry is complete at a level higher than 90\% down to $K_\mathrm{s}\approx$17.5, although this is an
average value because the completeness inevitably varies with crowding conditions in the wide area under study. The completeness
rapidly fades at fainter magnitudes, dropping below 50\% at $K_\mathrm{s}\approx$18.5.
\begin{figure}
\begin{center}
\includegraphics[width=9.5cm]{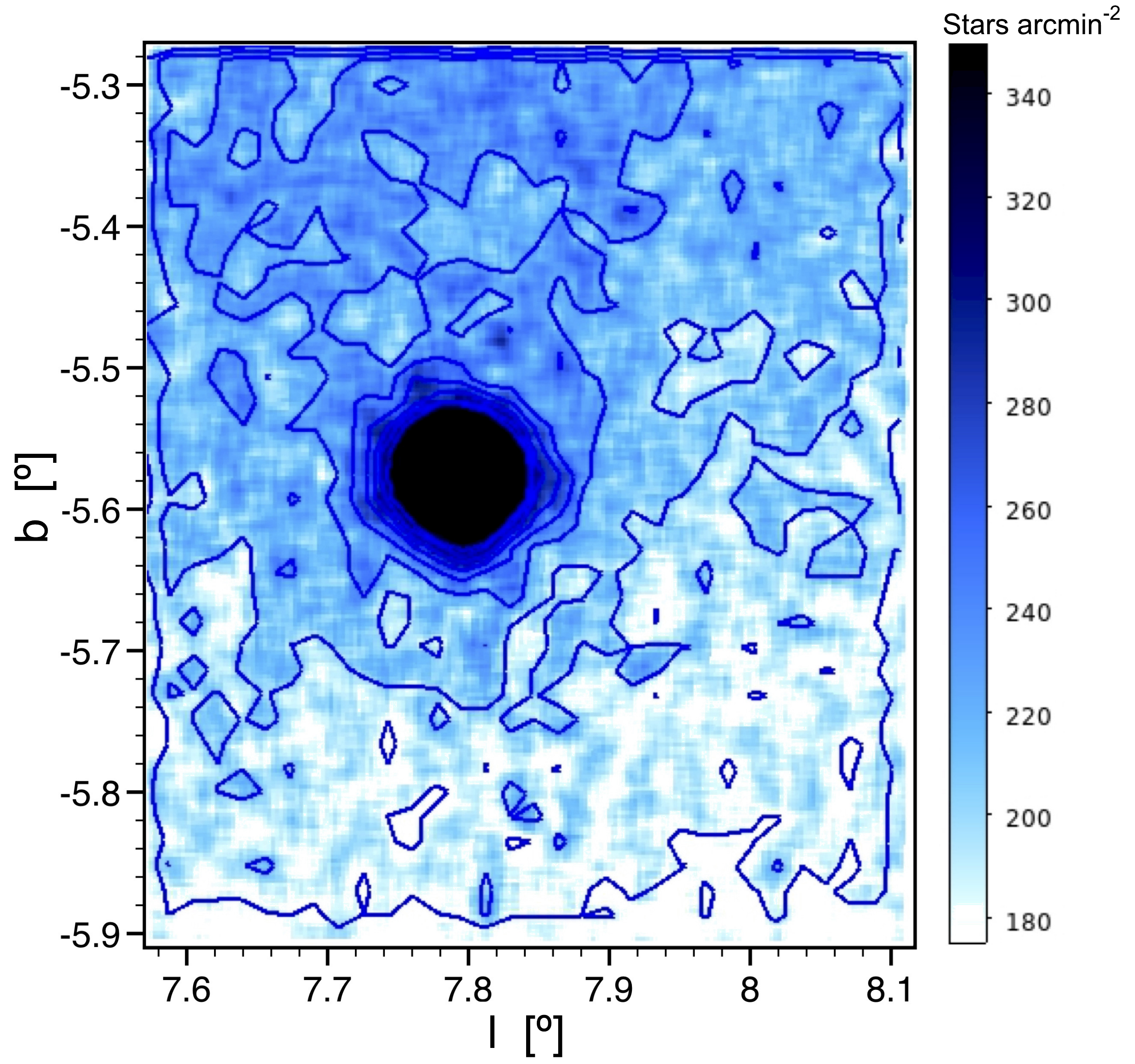}
\caption{Density map of the sources with $K_\mathrm{s}<17.5$ detected in the VVV frames. The density values in the blue-scale
bar are in units of sources per arcmin$^{2}$. The contour levels are shown at steps of 25 sources arcmin$^{-2}$ from the minimum
average background.}
\label{f_denmapgen}
\end{center}
\end{figure}

The density of the detected sources is shown in Fig.~\ref{f_denmapgen}. In the figure the cluster field is mapped with
$17\arcsec\times17\arcsec$ pixels, further smoothed with a $3\times3$ pixel boxcar. The map is relatively smooth, with a gentle
gradient increasing with the negative Galactic latitude (i.e. the stellar density increases slightly toward the Galactic plane).
The interstellar reddening is not constant in the field (see Sect.~\ref{ss_denmap} and Fig.~\ref{f_denmap}), but the stellar counts
show no feature that can be caused by variations of extinction. On the contrary, they increase with it toward the Galactic plane.
Hence, the stellar counts are negligibly affected by differential reddening.

\subsection{Spectroscopic data}
\label{ss_dataspec}

\begin{figure}
\begin{center}
\includegraphics[width=9.cm]{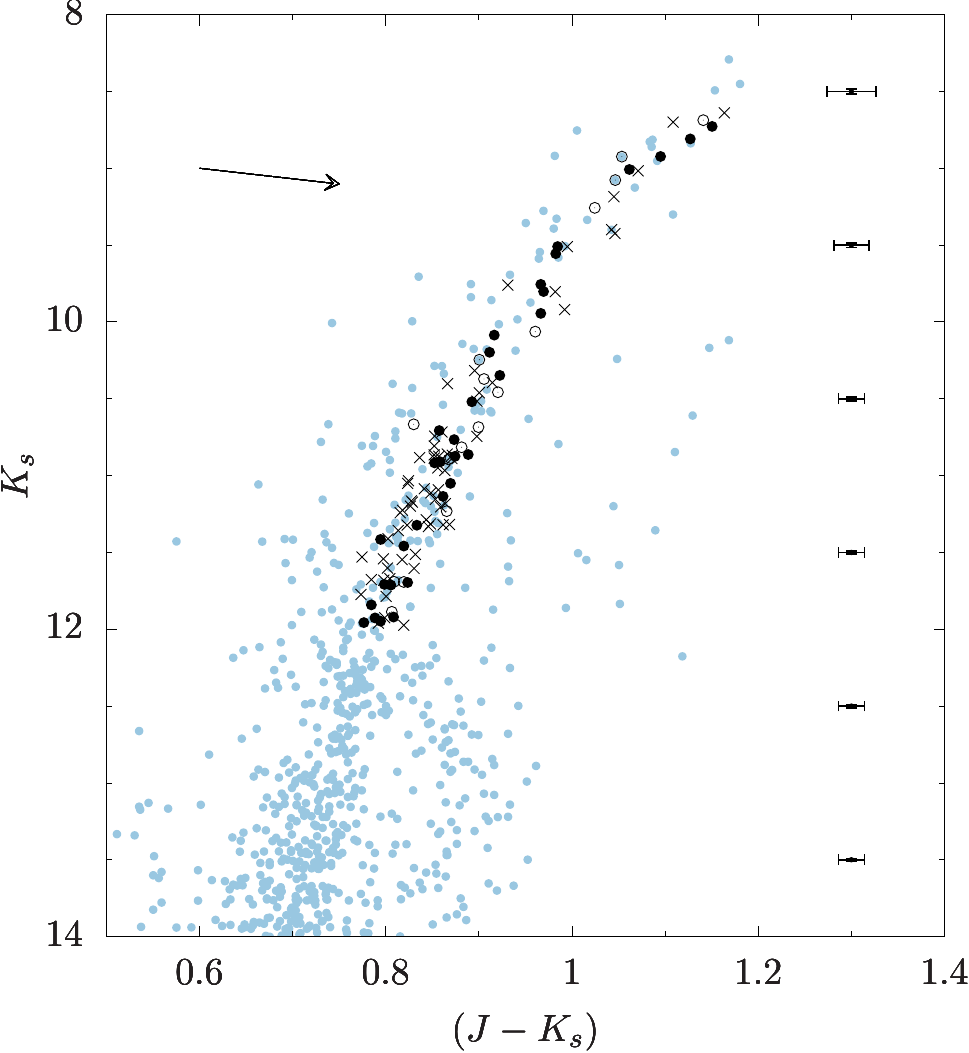}
\caption{Position of the spectroscopic targets in the VVV colour-magnitude diagram. The sources detected within 2$\arcmin$ of the
cluster centre are overplotted in light blue. Solid dots, open symbols, and crosses indicate the stars classified as cluster
members, additional member candidates, and field stars, respectively. Average errors as a function of magnitudes are plotted
on the right, and the arrow shows the direction of interstellar reddening.}
\label{f_CMDspec}
\end{center}
\end{figure}

The spectra of 113 stars in the field of M\,28 were collected in service mode with the ESO high-resolution multi-fibre spectrograph
FLAMES/GIRAFFE \citep{Pasquini02} mounted on the VLT-UT2. The employed grating LR8 returned spectra in the range 8180--9365~\AA,
where the CaT spectral feature (a triplet of calcium lines at 8498, 8542, and 8662~\AA) is found, with resolution R=6500. All
the spectra were collected with a single 900s exposure with one fibre configuration. The targets were selected by VVV photometry,
to span the upper cluster RGB in the range $K_\mathrm{s}$=8.5--12. Priority was given to the stars in the inner cluster regions;
however, due to fibre positioning limitations, objects up to $12\farcm5$ from the centre were observed. Their position in the cluster
colour-magnitude diagram (CMD) is shown in Fig.~\ref{f_CMDspec}, while the fibre IDs, coordinates, and photometric data are given in
the first five columns of Table~\ref{t_targ}.

The data were pre-reduced (bias-subtracted and flat-fielded), extracted, and wavelength-calibrated with the dedicated
pipeline\footnote{http://www.eso.org/sci/software/pipelines/}. The wavelength calibration solution was based on the lamp spectra
collected simultaneously during observations with five dedicated fibres. The spectra were extracted with an optimum algorithm
\citep{Horne86}, and the average of 11 fibres allocated to the sky background was subtracted. The spectra were eventually
normalised by fitting a low-order polynomial to the continuum, avoiding the three strong \ion{Ca}{ii} lines falling in the spectral
range. The signal-to-noise ratio of the final spectra varied from $\sim$75 for the faintest targets ($K_\mathrm{s}\approx$12) to
about 300 for the brightest objects ($K_\mathrm{s}\approx$8.6).


\section{Infrared photometry}
\label{s_phot}

\subsection{Radial density profile}
\label{ss_raddens}

Our analysis starts with the study of the spatial distribution of cluster stars because the definition of the cluster area in
the wide field under analysis will often be needed throughout our work. Several studies have recently found that catalogue values
of GC centres are typically discrepant by at least several arcseconds \citep[e.g.][]{Goldsbury10,Watkins15}.

To determine the coordinates of the cluster centre, we calculated the average position of the stars within 3$\arcmin$ from
the centre given by \citet{Noyola06}, and the result was used as input of a new iteration of the algorithm. The convergence was
reached in three iterations, when the new centre differed from the input value by less than one pixel ($0\farcs$34). We
restricted the analysis to objects brighter than $K_\mathrm{s}=$16 to reduce the effects of incompleteness and the contamination
of the Galactic background. The routine implicitly assumes that the field counts are homogeneous, else the result could drift from
the cluster centre. However, the photometric cut of faintest sources (where the ratio of cluster to field stars is lowest) and the small
area employed in the calculation prevented the small gradient seen in Fig.~\ref{f_denmapgen} from affecting the results. The
centre was found at RA=18:24:32.58, Dec=$-$24:52:13.6, less than 5$\arcsec$ from the definition of \citet{Noyola06}, and
$2\farcs1$ from that of \citet{Miocchi13}.

\begin{figure}
\begin{center}
\includegraphics[width=9.cm]{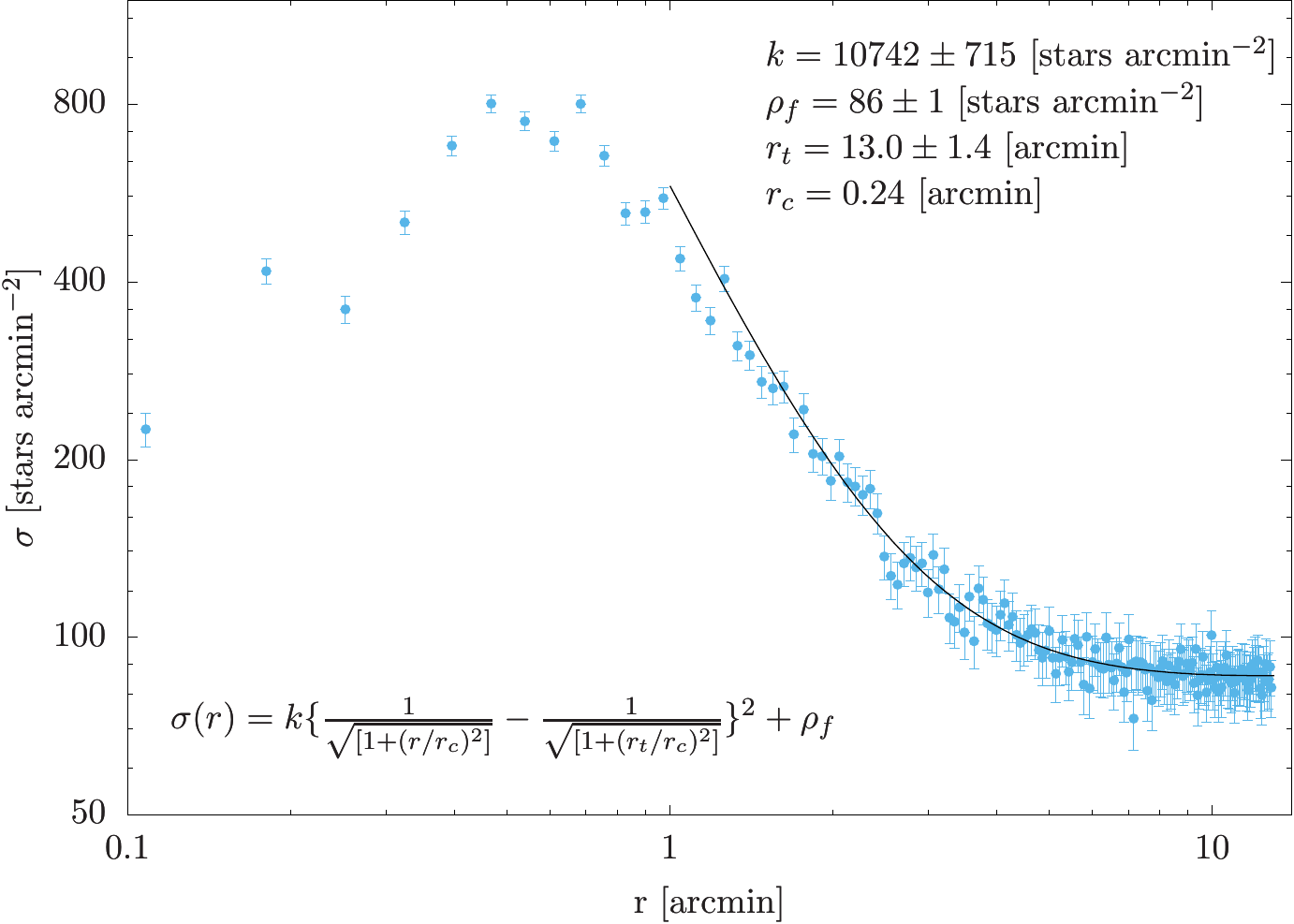}
\caption{Radial density profile of detected sources with $K_\mathrm{s}<16$. The curve shows the best-fit King profile in the range
$r=2-13\arcmin$.}
\label{f_radprof}
\end{center}
\end{figure}

The centre previously defined was used to derive the radial density profile. To this end, we divided the area up to 13$\arcmin$
from the cluster centre in 180 annuli centred on it, with width $r_\mathrm{out}-r_\mathrm{in}=4\farcs34$. We then calculated
the density of the sources with $K_\mathrm{s}<16.5$ detected in each annulus. The result is shown in Fig.~\ref{f_radprof}. The
stellar counts rapidly decline outward to a constant value of $\sim$86 stars per arcmin$^2$ at $r>11\arcmin$, but they are hardly
distinguishable from the field density already at $r\sim4-5\arcmin$. The star counts are severely affected by crowding in the
central region, and the density stops increasing inward at $r\sim 0\farcm5$. This incompleteness prevents us from studying the
density profile in the inner area. Hence, we fit a \citet{King62} profile to the data at $r>2\arcmin$, fixing the core radius
$r_c=0\farcm24$ from \citet{Trager93}. We verified that the results are insensitive to this assumption, mainly because our fit
only deals with the outer regions and the bin width is narrower than $r_c$ itself. As a consequence, the results were unchanged
when assuming alternative values from the literature \citep[$0\farcm26$ or $0\farcm21$,][]{Chun15,Kerber18}. The best fit is
shown in Fig.~\ref{f_radprof}, and it returns $r_t=13\farcm0\pm1\farcm4$. The fit curve also shows that the stellar counts start
to be affected by incompleteness already at $r\approx1\farcm5$.

Previous measurements in the literature for M\,28 report a tidal radius of $11\farcm2$ \citep{Trager93} and $12\farcm0$
\citep{Chun15}. Our result is slightly larger than theirs, but compatible within the errors. However, it must be considered
that the non-uniform stellar background seen in Fig.~\ref{f_denmapgen}, and the presence of the tidal structures discussed in
Sect.~\ref{ss_denmap}, can easily lead to an overestimate of the tidal radius.

\subsection{Decontaminated colour-magnitude diagram}
\label{ss_cmd}

\begin{figure}
\begin{center}
\includegraphics[width=9.cm]{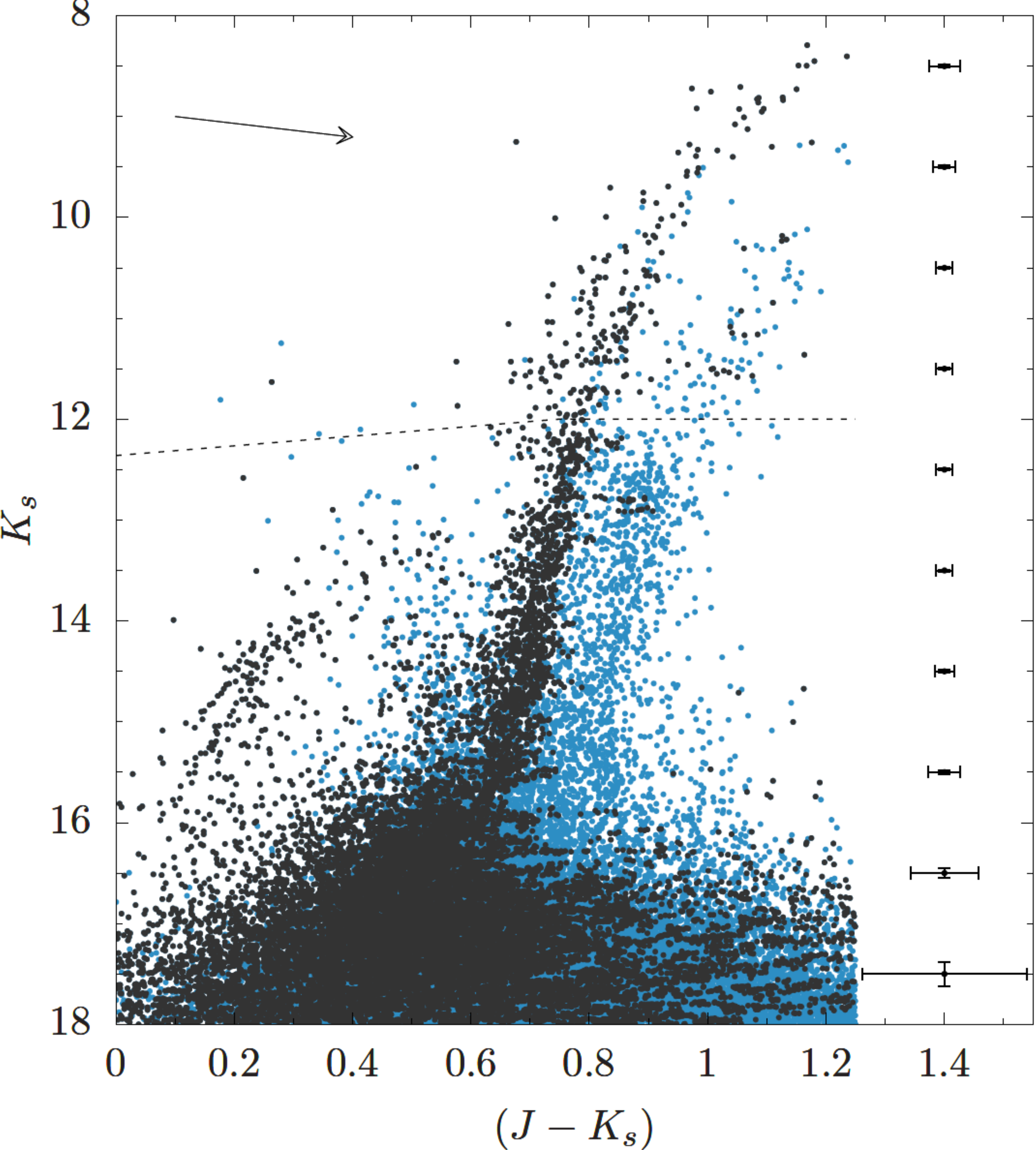}
\caption{Original (blue dots) and decontaminated (black dots) CMD of the cluster area ($r<4\arcmin$). The arrow and the error
bars are as in Fig.~\ref{f_CMDspec}. The dotted line shows the saturation limit in VVV frames.}
\label{f_decCMD}
\end{center}
\end{figure}

Figure~\ref{f_radprof} shows that although the cluster occupies a large area in the sky, its density rapidly approaches the field
background in the outer regions. For example, the annulus between $r=4\arcmin$ and $r=5\arcmin$ would statistically include about
400 cluster stars, but a field population six times larger. We therefore limit the cluster study to the area $r<4\arcmin$
(hereafter defined as `cluster area'), to maximise the contrast between cluster stars and field contamination. The field
contamination is relevant even in this area, where about 4300 field stars are expected out of a total of $\sim$9200 detections.
Hence, we defined a field control area as an annulus centred on the cluster, with inner radius $r_\mathrm{inn}=11\farcm5$ and
area equal to the cluster area. This region is still within the cluster tidal radius, but the expected cluster star density is so
low in it that its effect on the following analysis is negligible.

We adopted a procedure similar to that used in \citet{Moni11}, based on the method of \citet{Gallart03}, to clean the CMD of the
cluster area from the field contamination. The code scans the list of stars in the field area, and for each object it finds the
star in the cluster area with the smallest distance $d$ in the CMD, defined by
\begin{equation}
d=\sqrt{(\Delta K_\mathrm{s})^2+(k\times (\Delta (J-K_\mathrm{s}))^2},
\label{e_metric}
\end{equation}
where $k$ is an arbitrary coefficient weighting a difference in colour with respect to a difference in $K_\mathrm{s}$. A match is
obtained when $d$ is smaller than an arbitrary threshold $d_\mathrm{max}$, and the star is removed from the catalogue of cluster
sources. The parameters $d_\mathrm{max}$ and $k$ involved in the process are, however, rather arbitrary. We fixed
$d_\mathrm{max}$=0.3~magnitudes as in \citet{Gallart03} and \citet{Moni11}; this choice revealed a good compromise
between enough tolerance and the need to avoid associations between stars with very different photometry. \citet{Gallart03}
assumed $k$=7 in their optical photometry, while \citet{Moni11} extensively argued that in their VVV data a much smaller value
($k$=1--2) was the preferable choice. We therefore experimented with small values, and eventually adopted $k$=1 in our study.

\begin{figure}
\begin{center}
\includegraphics[width=9.cm]{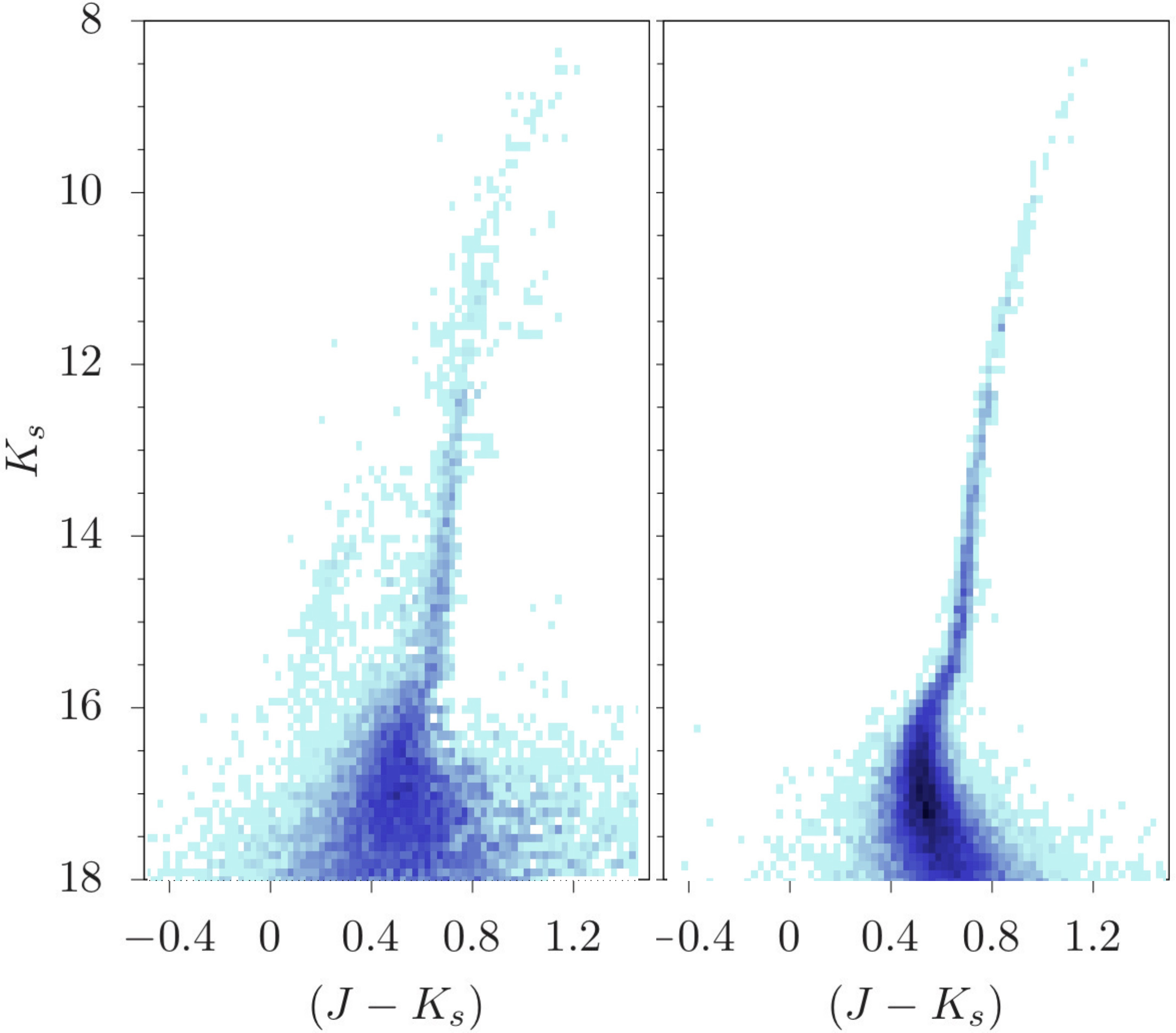}
\caption{Hess diagram of the decontaminated cluster CMD ({left panel}), and of a synthetic CMD calculated from the cluster
isochrone and the photometric errors ({right panel}).}
\label{f_mock}
\end{center}
\end{figure}

The decontaminated CMD is shown in Fig.~\ref{f_decCMD}, superimposed on all sources from the cluster area. All the cluster
sequences clearly emerge in the cleaned diagram, in particular the sub-giant (SGB) and asymptotic giant (AGB) branches,
and a well-populated blue HB. Most of the field stars have been removed by the decontamination procedure, in particular the
background RGB redder and fainter then the cluster sequence. A very slight residual contamination remains, observable in
particular in correspondence to the field red clump and its upper main sequence (MS). This is most likely a consequence of the
photometric incompleteness in the central cluster regions, which causes a small global under-detection of field stars. The few
objects found between the cluster HB and RGB, however, could be (at least in part) cluster blue stragglers rather than
field contaminants. This residual field contamination takes the shape of horizontal patterns in the faint red edge of
the CMD, due to our choice of $k=1$. This value implies that an offset in magnitude and in colour have the same weight
in Eq.~\ref{e_metric}, but the circles of equal $d$ become very elongated ellipses in a CMD that spans ten magnitudes in
$K_\mathrm{s}$ and only one in $(J-K_\mathrm{s})$.
 
Before the study of the decontaminated CMD, we felt it was worth analysing the differential reddening in the cluster field to
check to what extent it can affect the results. To this end, we generated a series of artificial CMDs by means of Monte Carlo
simulations, based on the cluster isochrone derived in Sect.~\ref{ss_isos} and the photometric errors as a function of
magnitude shown in Fig.~\ref{f_photerr}. The quantity of artificial stars at each magnitude was fixed to match the observed
luminosity function of the decontaminated CMD to take into account observational effects such as the catalogue incompleteness
in the inner regions. The results are shown in Fig.~\ref{f_mock}. The two panels look very similar, except for the
aforementioned residual contamination in the observed diagram, and the presence of observed HB and AGB stars, not accounted
for in the simulation. We calculated the width of the lower RGB in both the observed and synthetic diagrams in the magnitude
interval $K_\mathrm{s}=13.5-14.5$, assuming the difference as being entirely due to differential reddening in the cluster area. For
the observed CMD, a two-sigma clipping algorithm was applied to exclude from the calculation the cluster HB, and the residual
contamination by the bulge MS seen on the redder side of the cluster RGB at this magnitude range. The observed sequence is
slightly larger than the simulated one, and the quadratic difference of their widths indicates that the reddening variation
is about $\Delta E(J-K_\mathrm{s})\approx0.024$, equivalent to $\Delta E(B-V)\approx0.05$. This value is in line with the
reddening map shown in Fig.~\ref{f_denmap}. This result shows that the differential reddening is small in the cluster field,
and it will not be a major issue in the following analysis of the CMD.

\subsection{Isochrone fit}
\label{ss_isos}

\begin{figure}
\begin{center}
\includegraphics[width=9.cm]{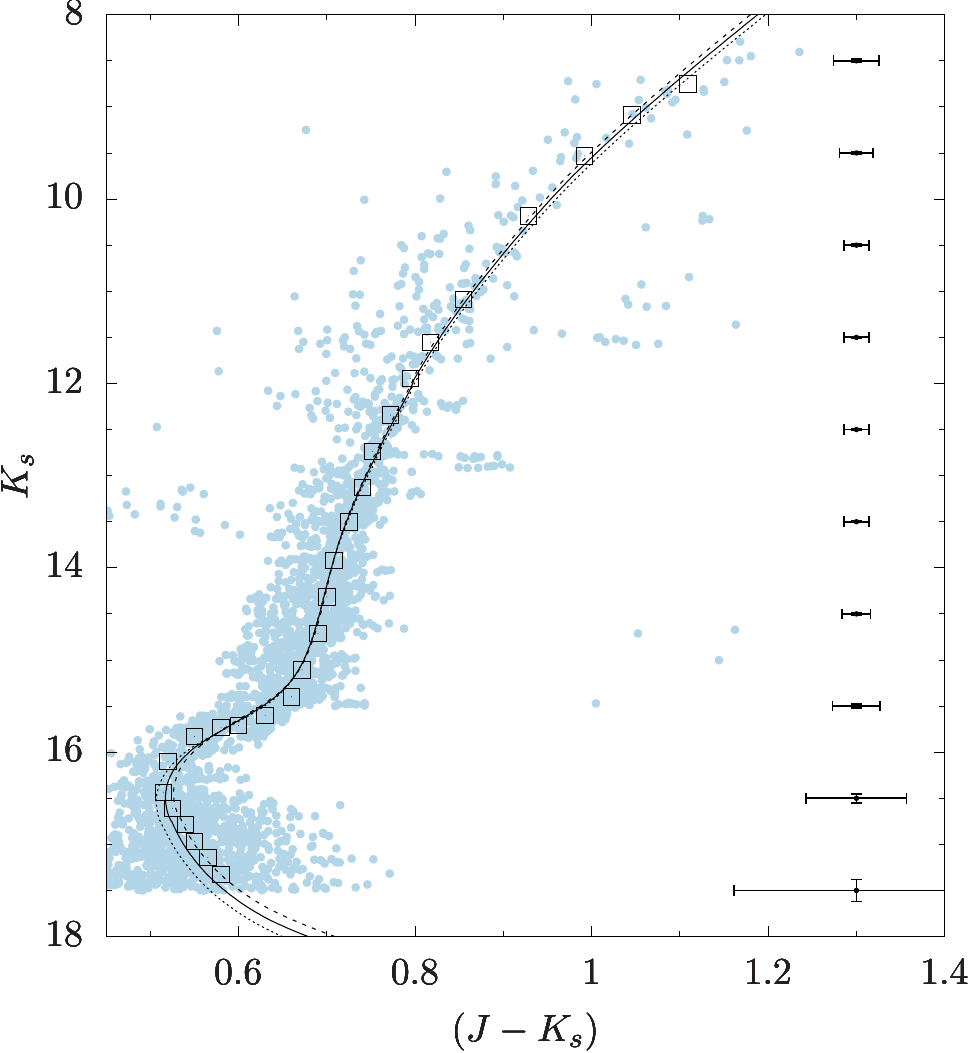}
\caption{Fiducial line of the cluster sequence (squares), and the best-fit DSED isochrones for $\tau$=12.5, 13.5, and 14.5~Gyr
(dotted, full, and dashed curve, respectively). The decontaminated cluster CMD is shown as light blue points.}
\label{f_isos}
\end{center}
\end{figure}

We used the decontaminated CMD to derive the cluster reddening, distance, and age by means of isochrone fitting. We first
defined the fiducial cluster sequence, with a fourth-order polynomial, fitting three sections of the CMD separately: the RGB
($K_\mathrm{s}<$15.5), the upper MS and turnoff region ($K_\mathrm{s}>$16), and the SGB intermediate between them. HB and AGB
stars, along with the residual contamination from the field upper MS at about $K_\mathrm{s}\approx$15,
$(J-K_\mathrm{s})\approx0.55$, were excluded from the fit to avoid spurious results. Nevertheless, the upper AGB merges with the
RGB at its brightest end, and it was probably not fully removed. In addition, we applied a reduction algorithm in the turnoff and
upper MS part to enhance the visibility of the cluster sequence and facilitate the fit. This consisted in dividing the data in
random groups of ten stars of similar magnitude ($\Delta K_\mathrm{s}<0.02$), and substituting each group with a single point
with its mean colour and magnitude. The results of this reduction algorithm, along with the fiducial cluster sequence, are shown
in Fig.~\ref{f_isos}.

We used the isochrones from the Dartmouth Stellar Evolutionary Database \citep[DSED;][]{Dotter08} to derive the cluster
parameters. We adopted a set of ages from $\tau$=11 to 15~Gyr in steps of 0.5~Gyr, with fixed values
$\left[ \frac{\mathrm{Fe}}{\mathrm{H}}\right]=-1.3$ and $\left[ \frac{\mathrm{\alpha}}{\mathrm{Fe}}\right]=-0.4$
\citepalias[][]{Villanova17}. \citet{Testa01} found evidence for a canonical helium abundance in M\,28, and \citet{Kerber18}
showed that even a mild helium enhancement of $\Delta Y=+0.05$ is not compatible with the mean magnitude of its RR~Lyrae stars.
Hence, we adopted a normal helium abundance ($Y$=0.2477) in our study.

The SGB and lower RGB ($K_\mathrm{s}$=13--16) are the best-defined sequences in the diagram. Hence, we used them as references to
adjust the cluster reddening $E(J-K_\mathrm{s})$ and distance modulus $(m-M)_{K_\mathrm{s}}$ for each isochrone independently,
leaving the upper RGB and the turnoff region as diagnostics for $\tau$.
All the isochrones returned the same value of reddening $E(J-K_\mathrm{s})=0.20\pm0.02$,
where the error was deduced by the width of the lower RGB, which is the dominant source of uncertainty when defining the
fiducial cluster sequence and the best-fit isochrone. Assuming a standard reddening law and the absorption relations of
\citet{Cardelli89}, this translates into $E(B-V)=0.38\pm0.04$. This result is slightly larger than the value $E(B-V)=0.35$ or
0.36 given by the \citet{Schlegel98} maps corrected as prescribed by \citet{Bonifacio00} or \citet{Schlafly11}, respectively,
but it agrees with older literature estimates, which are in the range $E(B-V)=0.37-0.40$ \citep{Webbink85,Zinn85,Reed88}. More
recent measurements preferred even higher values, $E(B-V)=$0.42-0.44 \citep{Davidge96,Testa01,Kerber18}. The assumption of a
standard reddening law might be weak in this region of the sky, but we do not detect any systematic difference between the
estimates based on IR photometry \citep[][and ours]{Testa01}, and those based on optical data.

The best-fit distance modulus decreases with age, at the approximate pace of about 0.1~magnitudes per Gyr, from
$(m-M)_{K_\mathrm{s}}=14.0$ at $\tau$=10.5~Gyr to $(m-M)_{K_\mathrm{s}}=13.5$ at $\tau$=14.5~Gyr. The best-fit isochrones for
cluster age $\tau$=12.5, 13.5, and 14.5~Gyr are shown in Fig.~\ref{f_isos}. By construction, they all coincide on the lower RGB,
where they were forced to match the cluster sequence to derive the best-fit reddening and distance modulus. They all fit the RGB
rather well up to $K_\mathrm{s}$=12, but at brighter magnitude the results are uncertain due to the paucity of stars and the
probable contamination of the derived cluster sequence by AGB objects. Some stars in the upper RGB, however, could be variable,
which adds additional uncertainty to the fit of the brightest sequence. On the fainter end of the diagram
($K_\mathrm{s}>$15.5), on the other hand, the curves present different behaviours. Isochrones younger than 13~Gyr are definitely
too blue on the main sequence, while the oldest solutions with $\tau\geqslant$14~Gyr fail to reproduce the turnoff region at
$K_\mathrm{s}\approx$16. We therefore conclude that M\,28 must be a very ancient object, with age in the range $\tau$=13--14~Gyr.
This is older than the age found by \citet{Kerber18} with the same DSED isochrones (12.1$\pm1.0$~Gyr), although they found an older
age (14.1$\pm$1.0~Gyr) with BaSTI models \citep{Pietrinferni06}. Within this age range, the distance modulus is constrained to
$(m-M)_{K_\mathrm{s}}=13.6\pm0.1$, which translates into an absolute distance modulus of $(m-M)_0=13.5\pm0.1$, and a distance of
$d=5.01\pm0.23$~kpc, assuming a standard reddening law to derive $A_{K_\mathrm{s}}$ from $E(J-K_\mathrm{s})$. This assumption is
safe in this case because $A_{K_\mathrm{s}}$ is small (of the order of 0.1~magnitudes), and any local deviations from the
standard law produces a deviation one order of magnitude smaller than the error on $(m-M)_{K_\mathrm{s}}$. Our distance
estimate is very similar to the recent result of
\citet[][$(m-M)_0=$13.51-13.67, depending on the isochrone set used]{Kerber18}, based on Hubble Space Telescope data.

\subsection{Spectroscopic vs photometric stellar and cluster parameters}
\label{ss_specred}

The spectroscopically derived temperature, gravity, and metallicity values of cluster members can be used to derive the cluster
reddening and distance, avoiding the high degeneracy of multiple-parameter isochrone fits. Unfortunately, spectroscopic and
photometric parameters often differ, and the differences strongly increase at low metallicity
\citep[see e.g.][and references therein]{Sbordone15}. Here we check the consistency of the spectroscopic measurements of
\citetalias{Villanova17} with those that can be derived from our IR photometry.

\begin{figure}
\begin{center}
\includegraphics[width=9.cm]{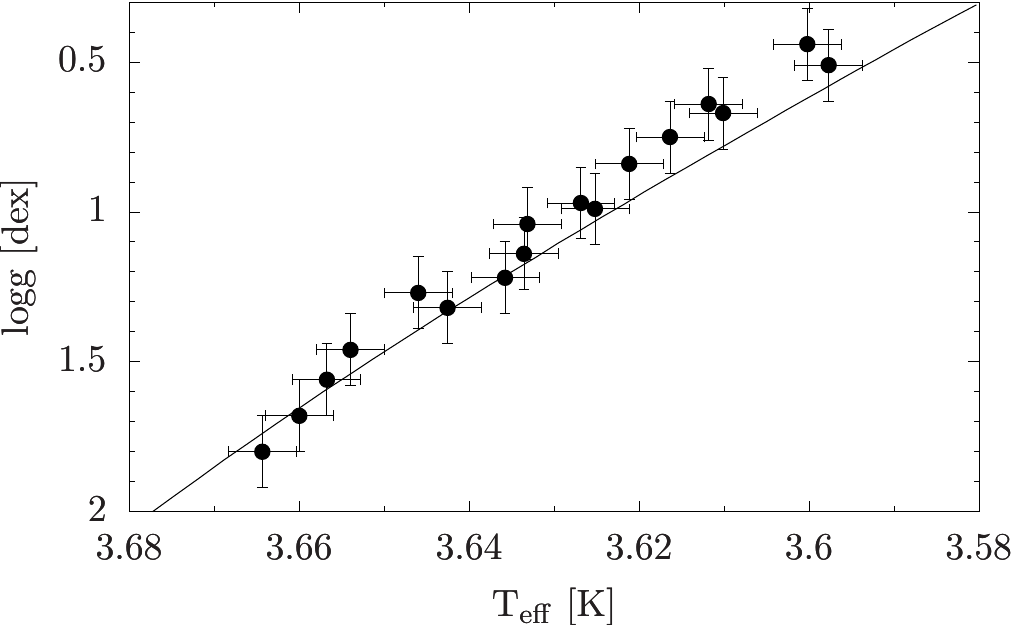}
\caption{Comparison between the \citetalias{Villanova17} results (filled dots) and DSED isochrones (full curves), in the
temperature-gravity plane. The errors shown are the internal uncertainties from \citetalias{Villanova17}.}
\label{f_V17}
\end{center}
\end{figure}

In Fig.~\ref{f_V17} we compare the \citetalias{Villanova17} results in the temperature-gravity plane with a DSED isochrone with
$\tau$=13.5~Gyr and $\left[ \frac{\mathrm{Fe}}{\mathrm{H}}\right]=-1.29$ \citepalias{Villanova17}. The position of stars in this
diagram is independent of reddening, distance, and age. We show only one isochrone in the figure because DSED curves with
$\tau$=12.5, 13.5, 14.5~Gyr are totally indistinguishable in the plot. Figure~\ref{f_V17} suggests a general agreement between the
\citetalias{Villanova17} and the DSED isochrones, but also the presence of a systematic, because the sample shows a drift to lower
gravities at decreasing temperatures. This might be a consequence of deviations from local thermodynamic equilibrium (LTE), as
non-LTE over-ionisation in the atmosphere of cool red giants can offset the spectroscopic gravities based on LTE ionisation
equilibrium of iron lines \citep[see e.g.][]{Sitnova15,Mashonkina16}.

We transformed the spectroscopic effective temperature of the 17 targets of \citetalias{Villanova17} into a theoretical
colour $(J-K_\mathrm{s})_0$ by means of the temperature-colour relation of \citet{Alonso99}. These relations are valid for
$(J-K_\mathrm{s})<1$, while the observed colour of our reddest targets is $(J-K_\mathrm{s})\approx1.1$. However, taking into account
the reddening obtained in Sect.~\ref{ss_isos}, their intrinsic colour is still within the range of validity of the \citet{Alonso99}
relations. The $(J-K)$ colour in the TCS system used by \citet{Alonso99} was transformed to the 2MASS system by means of the
2MASS\footnote{http://www.astro.caltech.edu/\~{}jmc/2mass/v3/transformations/} and \citet{Alonso98} equations. From the difference
with the observed VVV colour, we obtained an estimate of the reddening for each star. The results span the range
$E(J-K_\mathrm{s})=0.12-0.22$, with average and standard deviation of $E(J-K_\mathrm{s})=0.17\pm0.03$. Under the assumption of a
standard absorption law, this translates into $E(B-V)=0.33\pm0.04$. We do not see a clear gradient or pattern in the spatial
distribution of $E(B-V)$ of the 17 stars, but the average value of the western half of the sample is 0.03~magnitudes more
reddened than the eastern half. This difference is negligible compared to the errors, but in line with the \citet{Schlegel98} map
that shows a tiny gradient of $\Delta E(B-V)=0.02$ increasing toward the west.

The spectroscopically derived reddening is lower than previous measurements in the literature, but still compatible within
$1\sigma$ with our photometric estimate. To match our reddening estimate based on the isochrone fit, the \citetalias{Villanova17}
temperatures should be cooler by $\sim80$~K. A correction of all temperatures by this quantity would drift the observed points in
Fig.~\ref{f_V17} away from the theoretical isochrone, and an additional correction of gravities by $+0.20$~dex would be needed to
recover the agreement in this plot. The errors quoted by \citetalias{Villanova17} are smaller than these systematics by about a
factor of two, but they represent only internal uncertainties.

If absolute magnitudes are derived from the isochrone assuming the \citetalias{Villanova17} parameters, the distance modulus is
overestimated by $+0.8$~magnitudes with respect to the value obtained from the isochrone fit (Sect.~\ref{ss_isos}), and the resulting
distance d$\approx$7.5~kpc is larger by 50\%. We find that an offset of $+160$~K in temperature and $+0.35$~dex in gravity should
be applied to the \citetalias{Villanova17} measurements to force a match with isochrone fit reddening and distance modulus, while
still keeping the agreement with the theoretical isochrone found in Fig.~\ref{f_V17}.

To further compare photometric and spectroscopic temperatures, we elaborated a method to estimate temperatures from photometric
colours that can account for differential reddening and a non-standard reddening law. We therefore projected the position of the
17 \citetalias{Villanova17} targets in the Two Colour Diagram (TCD) onto the same DSED isochrone used in Fig.~\ref{f_V17}
($\tau$=13.5~Gyr and $\left[ \frac{\mathrm{Fe}}{\mathrm{H}}\right]=-1.29$). We adopted the
$(J-K_\mathrm{s})-(G_\mathrm{bp}-K_\mathrm{s})$ plane, where $G_\mathrm{bp}$ is the bluest Gaia band, because $(G-K_\mathrm{s})$
is highly reddening-dependent while $(J-K_\mathrm{s})$ is poorly affected by it. As a consequence, this combination returns an
almost vertical reddening line, maximising its angle with the DSED isochrone and thus minimising the uncertainties of the
projection procedure.

To avoid constraining the reddening law, we did not fix the slope of the reddening line in the TCD. Instead, we first corrected
the $(J-K_\mathrm{s})$ colours by the average cluster reddening $E(J-K_\mathrm{s})=0.20$ previously determined, and connected each
star to the point on the isochrone at its de-reddened $(J-K_\mathrm{s})_0$. This calculation returned a different slope for each
star; the values were then averaged to obtain our best estimate of the reddening line slope. This value was eventually kept fixed
and used to project the targets onto the isochrone, and the intersection provided the photometric temperature and gravity of each
star. This procedure thus assumes a unique (but not necessarily standard) reddening law, but it allows different reddening for
each star because the length of the projection may vary from object to object.

We estimated the errors on the derived temperature and gravities calculating how these parameters were affected by the variation
of one input quantity. We assumed a change of 1~Gyr and 0.05~dex respectively for age and metallicity, the VVV photometric
errors for magnitudes and colours, our 1$\sigma$ uncertainty for $E(J-K_\mathrm{s})$; instead, for the slope of the reddening
line we adopted the error-on-the-mean obtained during the estimate of this quantity. These uncertainties were then quadratically
summed, obtaining an error of $\sim$100~K for T$_\mathrm{eff}$ and $\sim$0.2~dex for $\log{g}$.

The absolute magnitude in all bands can also be obtained from the projection on the isochrone. We thus derive
$E(J-K_\mathrm{s})=0.20\pm0.01$ and $(m-M)_{K_\mathrm{s}}=13.5\pm0.2$ as the mean reddening and distance modulus for the
17 sample stars. These values perfectly match those obtained by means of the isochrone fit in the CMD (Sect.~\ref{ss_isos}).

The photometric temperatures and gravities are systematically higher than the spectroscopic estimates of \citetalias{Villanova17},
with mean differences of $\Delta\mathrm{T_{eff}}=167\pm45$~K and $\Delta\log{g}=0.39\pm0.09$. These values are essentially
identical to those previously derived by comparison with cluster parameters derived from isochrone fit in the CMD. The standard
deviation of the differences indicate that the external errors of the \citetalias{Villanova17} measurements are approximately
of the same order of magnitude as the uncertainties of our photometric estimates. In conclusion, the spectroscopic stellar
parameters are much lower than the photometric values, and they lead to an underestimate of cluster reddening and a large
overestimate of distance.

\begin{figure}
\begin{center}
\includegraphics[width=9.cm]{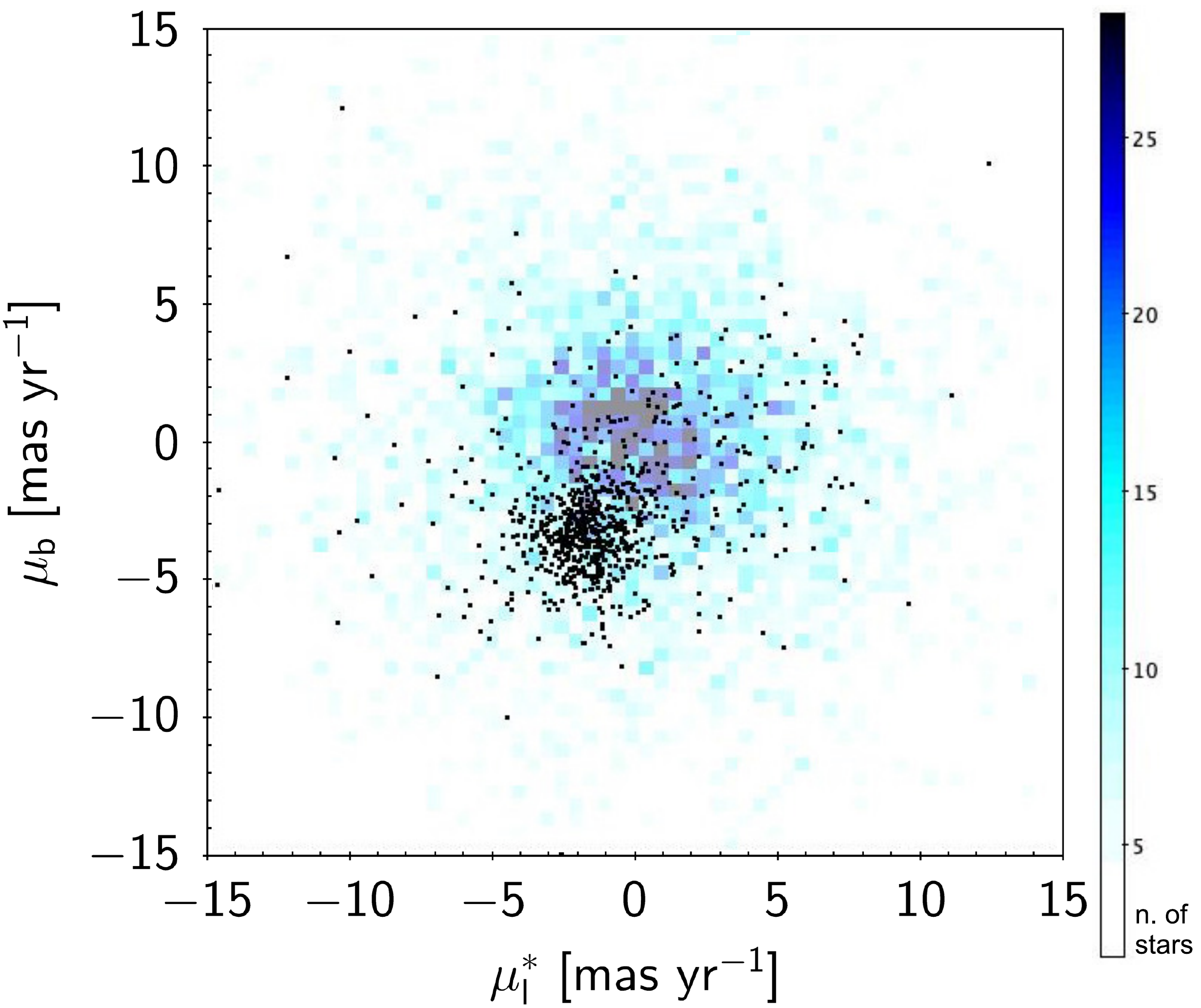}
\caption{VVV relative proper motions of stars in the cluster sequence (black dots), superimposed on the distribution of field
stars.}
\label{f_pm}
\end{center}
\end{figure}

\subsection{Proper motions}
\label{ss_pm}

The temporal interval spanned by VVV observations is large enough to measure stellar proper motions (PMs). We used 76 VVV epochs
acquired by the survey between 2010 and 2015, to derive the PM of all VVV sources within 6$\arcmin$ of the cluster centre. To
this end, we employed the same procedure described in detail in \citet{Contreras17}, which returns values relative to the mean
motion of bulge RGB stars in Galactic coordinates $(\mu_l,\mu_b)$. Whenever needed during the analysis, we used the formulae
from \citet{Poleski13} to transform $(\mu_l^*,\mu_b)$ to the equatorial components $(\mu_\alpha^*,\mu_\delta)$, where
$\mu_l^*=\mu_l\cdot\cos b$ and $\mu_\alpha^*=\mu_\alpha\cdot\cos\delta$. Our final catalogue comprised more than 64000 sources down
to $K_\mathrm{s}\sim18.3$.

The computed PMs show the presence of two populations: a wide distribution made up of bulge stars centred at
$(\mu_l^*,\mu_b)\approx(0,0),$ as expected, and a small clump offset from the main distribution by $\sim$1-2~mas~year$^{-1}$ in
both coordinates. This is a clear signature of the different kinematics of the cluster with respect to the Galactic bulge. This
conclusion is demonstrated in the vector point diagram (VPD; see Fig.~\ref{f_pm}), where we show the PM distribution of stars
located along the cluster and field sequences observed in Fig.~\ref{f_decCMD}.

To derive the mean relative PM of the cluster, we performed a visual selection of cluster stars using both the CMD and the VPD.
The procedure returned about 1000 likely members with $K_\mathrm{s}<15$ close ($\Delta(J-K_\mathrm{s})\approx\pm0.02$) to
the cluster red RGB. After cleaning with a 2$\sigma$ clipping algorithm to remove residual field contamination, we obtain the
average cluster value of $(\mu_l^*,\mu_b)=(-1.54\pm0.05,-3.32\pm0.04)$~mas~yr$^{-1}$. The quoted errors are the statistical
errors-on-the-mean. To derive the absolute values, we matched the list of field RGB stars that have been used as reference stars
to derive relative PMs with the GAIA DR2 database \citep{Gaia16,Gaia18}. We restricted the reference star sample to those stars
showing relatively low PM errors ($<$2~mas~yr$^{-1}$). After a 2$\sigma$ clipping algorithm to clean the list from outlier
measurements, the matched list comprised $\sim$400 objects. The comparison with GAIA measurements revealed a zero-point absolute
offset of our relative PMs of $\Delta(\mu_l^*,\mu_b)\approx(-6.21\pm0.06,-0.29\pm0.06)$~mas~yr$^{-1}$. After applying this
correction to the cluster value, the derived absolute PM for M\,28 resulted
$(\mu_\alpha^*,\mu_\delta)=(-0.35\pm0.08,-8.54\pm0.08)$~mas~yr$^{-1}$ in equatorial components. Our result is in good agreement
with the latest results of GAIA \citep{Gaia18b}, and the later estimate of \citet{Vasiliev18}, while there are some small
differences with previous values from the literature, as shown in Table~\ref{t_pm}.

\begin{table}[t]
\begin{center}
\caption{Literature estimates of the cluster proper motion.}
\label{t_pm}
\begin{tabular}{r l l}
\hline
\hline
$\mu_\alpha^*$ & $\mu_\delta$ & Reference \\
mas~yr$^{-1}$ & mas~yr$^{-1}$ & \\
\hline
$0.30\pm0.50$ & $-3.40\pm0.90$ & \citet{Cudworth93} \\
$0.63\pm0.67$ & $-8.46\pm0.67$ & \citet{Dinescu13} \\
$-0.81\pm0.18$ & $-6.85\pm0.18$ & \citet{Chemel18} \\
$-0.42\pm0.01$ & $-8.80\pm0.01$ & \citet{Gaia18b} \\
$-0.30\pm0.06$ & $-8.91\pm0.06$ & \citet{Vasiliev18} \\
$-0.35\pm0.08$ & $-8.54\pm0.08$ & This work \\
\hline
\end{tabular}
\end{center}
\end{table}

\subsection{Tidal structures}
\label{ss_denmap}

\begin{figure}[h!]
\begin{center}
\includegraphics[width=9.8cm]{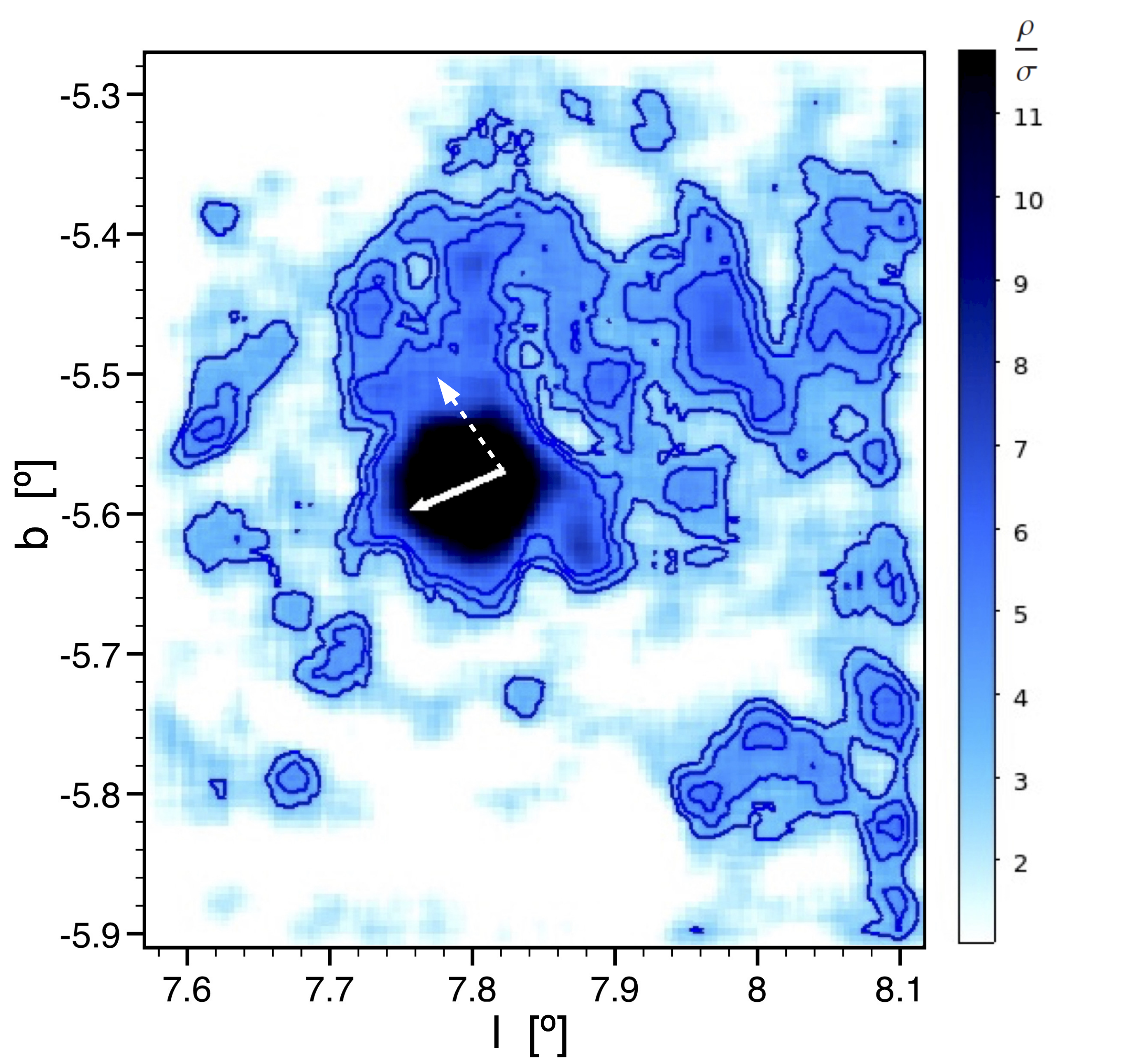}
\includegraphics[width=9.cm]{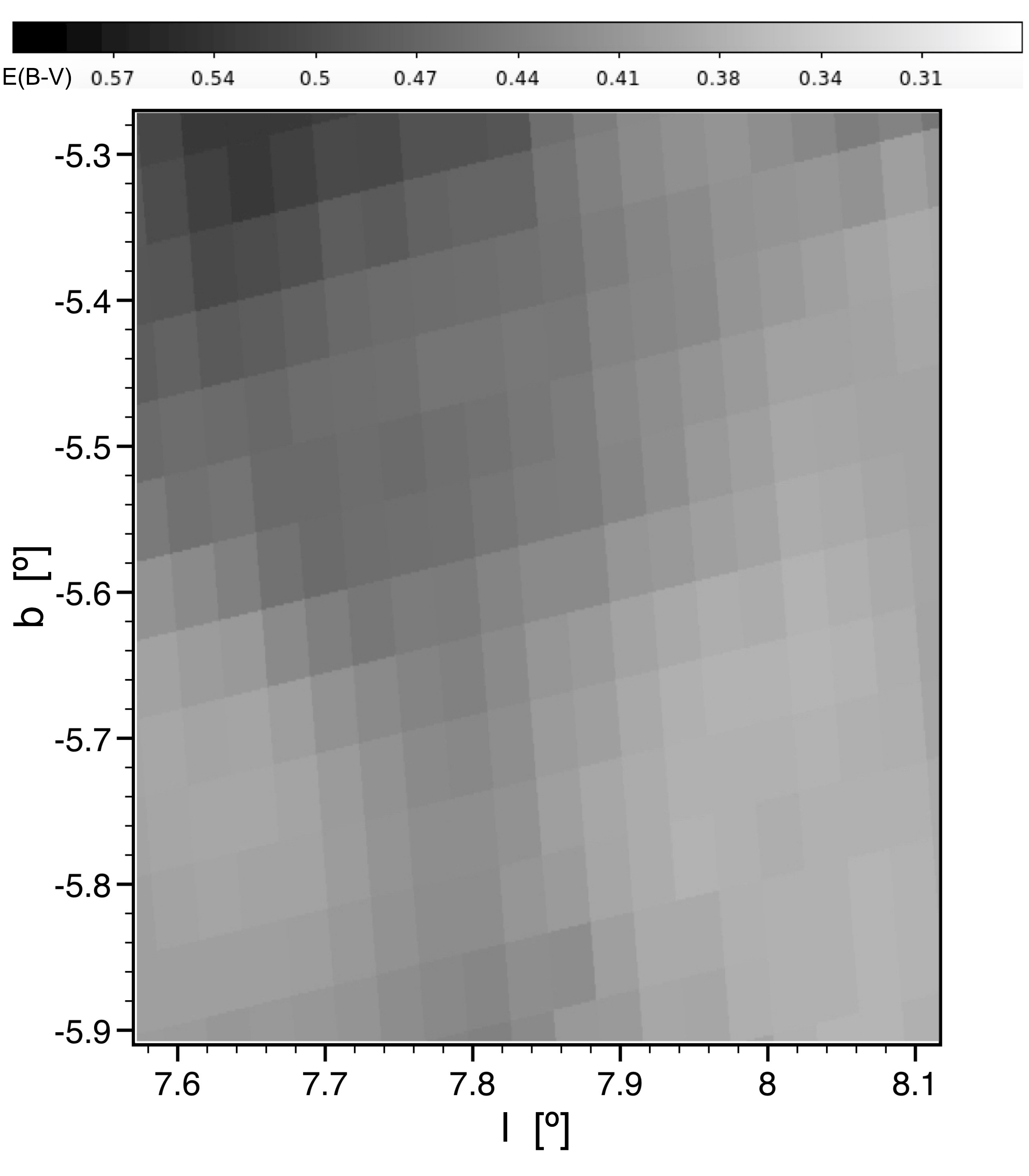}
\caption{{\it Upper panel}: Normalised background-subtracted density map of the cluster area, in units of the background noise
($\sigma$). Contour levels are shown at 3$\sigma$, 4$\sigma$, and 5$\sigma$. The full white arrow shows the direction of the
cluster proper motion, while the dashed arrow points toward the Galactic centre. {\it Lower panel}: \citet{Schlegel98}
reddening map of the same area as the upper panel. The reddening values in the greyscale bar are given in $E(B-V)$.}
\label{f_denmap}
\end{center}
\end{figure}

The wide field covered by VVV data is useful to study the possible presence of tidal structures around the cluster. To this
end, we generated the density map of all stars with $K_\mathrm{s}<17.5$ located within $\Delta(J-K_\mathrm{s})=\pm0.05$ from the
best-fit isochrone defined in Sect.~\ref{ss_isos}. This width was fixed to maximise the selection of cluster objects, taking into
account photometric errors and the reddening variations in the wide 40$\arcmin\times30\arcmin$ field under study. We also produced
a similar map including all stars with $K_\mathrm{s}<17.5$ to study the behaviour of the general field. We then normalised both
maps by the average stellar density calculated in the SW corner of the area, outside the cluster tidal radius where the field
density has its minimum. Then, the normalised map of the general field was subtracted from the cluster map to remove the trend of
the background field density.

The resulting map is shown in Fig.~\ref{f_denmap}, normalised in units of the background noise, along with the \citet{Schlegel98}
reddening map of the same area. The cluster is actually doughnut-shaped in the map due to the decrease of completeness of our
catalogue in the central regions (see Sect.~\ref{ss_raddens} and Fig.~\ref{f_radprof}), but this is not clear in the figure because
the saturation limit of the colour scale was set at a lower level to observe the faint structures far from the centre. This also
happened in Fig.~\ref{f_denmapgen} (compare the colour scale in Fig.~\ref{f_denmapgen} with the stellar counts in
Fig.~\ref{f_radprof}). Outside the inner dense regions, we detect a series of structures around the cluster. The most prominent
feature is a long clumpy stripe in the direction opposite to the cluster motion. In addition, a tail extending toward the Galactic
centre is also visible. A leading arm in the direction of the cluster motion is not found, while a reduced structure toward the
Galactic anti-centre might be present, but more distant from the cluster centre than the other features. These structures do not
follow the trend of the reddening map, nor the general gradient increasing with Galactic latitude observed in Fig.~\ref{f_denmapgen}.

\citet{Kundu19} studied RRLyrae extensions in 56 GCs, finding extra-tidal RR Lyrae in 20\% of them, but they did not detect such
stars in the area of M\,28. However, \citet{Chun12,Chun15} claimed the detection of two tidal tails extending toward the Galactic
centre and anti-centre, and a clumpy structure in the direction opposite to the cluster PM. All these structures are visible at a
$\sim2\sigma$ level in their maps \citep[see Fig. 6 of][]{Chun15}. Their work maps an area much larger than our, and
the tail toward the Galactic anti-centre is clearly visible only beyond the cluster tidal radius. Our results closely reproduce
theirs, confirming the existence of these tidal features.


\section{CaT spectroscopy}
\label{s_spec}

\subsection{Radial velocities}
\label{ss_rvs}

The radial velocity (RV) of target stars was measured cross-correlating \citep{Simkin74,Tonry79} the three CaT lines with
synthetic templates drawn from the library of \citet{Coelho05} with [Fe/H]=$-$1.5, [$\alpha$/Fe]=+0.4, and temperature and
gravity varying with the position of the star on the cluster RGB. The cross-correlation was performed with the
IRAF\footnote{IRAF is distributed by the National Optical Astronomy Observatories, which are operated by the Association of
Universities for Research in Astronomy, Inc., under cooperative agreement with the National Science Foundation.} task
{\it fxcor}. The measurements were reduced to heliocentric velocities, and the results are given in the sixth column of
Table~\ref{t_targ}. Our sample has five stars in common with \citetalias{Villanova17}, and the differences (in the sense
ours$-$theirs) are small, within $-$2.1 and +0.2~km~s$^{-1}$. However, our RV is smaller than that of
\citetalias{Villanova17} for four of these five objects, and the mean difference is $-$0.8~km~s$^{-1}$, which suggests the
presence of a tiny zero-point offset between the two measurement sets. This difference is of the order of our errors. We also
have four targets in common with \citetalias[]{Rutledge97}. The differences in this case are huge, with
\citetalias{Rutledge97} RVs being higher by 20-40~km~s$^{-1}$. The average difference is $+29$~km~s$^{-1}$, which is similar
to the difference between our estimate of the cluster RV and theirs (see Table~\ref{t_RVlit}).

\begin{figure}
\begin{center}
\includegraphics[width=9.cm]{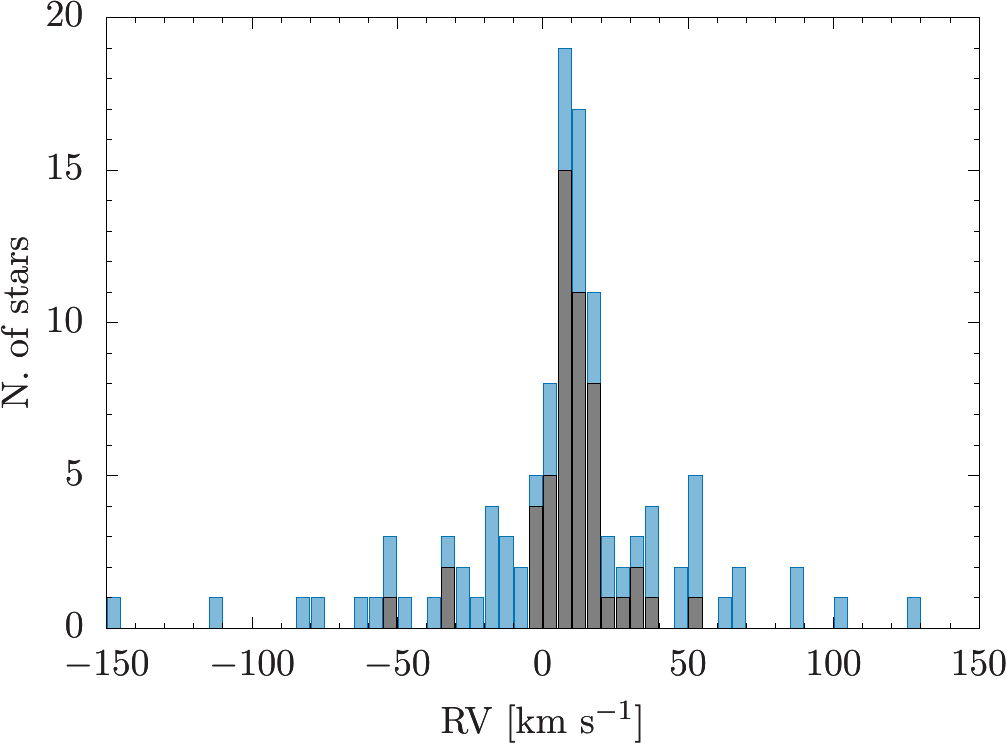}
\caption{Histogram of RV distribution of all spectroscopic targets (blue histogram), and targets within $4\arcmin$ from the
 cluster centre (black histogram).}
\label{f_histoRV}
\end{center}
\end{figure}

\begin{figure}
\begin{center}
\includegraphics[width=9.cm]{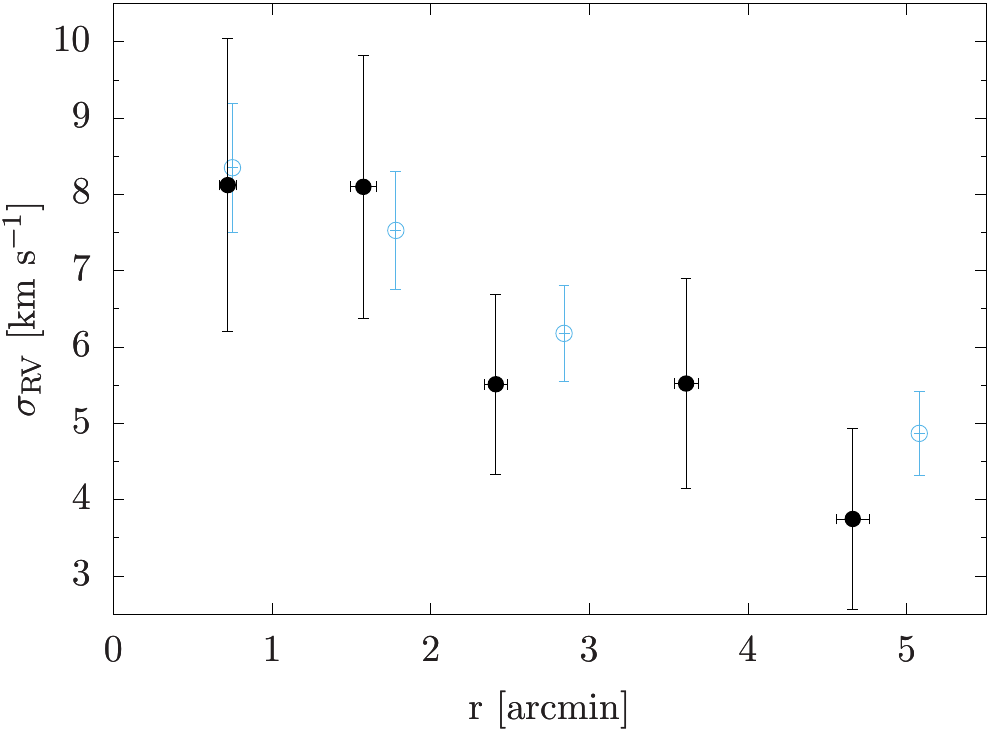}
\caption{Radial profile of the RV dispersion deduced from our sample. The big black dots are obtained binning our sample in steps of
$0\farcm1$. The blue symbols show the results of \citet{Baumgardt19}.}
\label{f_RVdisp}
\end{center}
\end{figure}

The RV distribution of all the targets is shown in Fig.~\ref{f_histoRV} as a grey histogram. A peak at RV$\approx$10~km~s$^{-1}$
is evident, but the measurements are distributed in a wide range, indicating a relevant contamination by field stars. To reduce
it and derive a more reliable cluster RV, we therefore analysed only the stars within $r<5\arcmin$ of the centre,
where the cluster dominates over the field density (see Fig.~\ref{f_radprof}). The RV distribution of this sub-sample is shown in
Fig.~\ref{f_histoRV} as a black histogram. Only seven outliers with $\vert$RV$\vert>$30~km~s$^{-1}$ are left, and they are
all found at the faintest end of the magnitude distribution ($K_\mathrm{s}>11$). This is not surprising as the field contamination
is expected to increase in the lower section of the RGB. We eventually obtained the cluster mean RV and dispersion fitting the
probability plot \citep{Hamaker78,Lutz92} of this $r<5\arcmin$ sample, excluding these six targets \citep[see][for more details
on the fitting procedure]{Moni12}. We thus find $\overline{\mathrm{RV}}=10.5\pm0.5$~km~s$^{-1}$ and
$\sigma_{\mathrm{RV}}=7.7\pm0.5$~km~s$^{-1}$, where the errors on these estimates was derived from the $\chi^2$ statistics of the
fit. This value of the velocity dispersion is lower than but still consistent with the $\sigma_{\mathrm{RV}}=9.4\pm1.5$~km~s$^{-1}$
found by \citetalias{Villanova17}. However, a unique definition of $\sigma_{\mathrm{RV}}$ from measurements of single stars is
not straightforward because the velocity dispersion varies in a GC with distance from the centre, and the observed targets are
distributed in a wide range of $r$. In Fig.\ref{f_RVdisp} we show the variation in $\sigma_{\mathrm{RV}}$ with radial distance
derived from our data. Our sample is not large enough to study the radial profile of the RV dispersion, but a decrease in
$\sigma_{\mathrm{RV}}$ with $r$ is clear. The trend derived by us is also very similar to the profile reported by
\citet{Baumgardt19}\footnote{see https://people.smp.uq.edu.au/HolgerBaumgardt/globular/}, shown in the same figure for comparison.
We find that the dispersion is roughly constant in the first 2$\arcmin$ from the centre, hence we limit the calculation to the
targets with $r<2\arcmin$, finding $\sigma_{\mathrm{RV,c}}=8.4\pm0.5$~km~s$^{-1}$ for the central value. This result, which should
still be regarded as a lower limit, matches the older estimate of \citet{Pryor93}, who proposed a central dispersion of
8.6$\pm$1.3~km~s$^{-1}$. We therefore confirm that the internal velocity dispersion is high in M\,28, indicating that this is a
massive cluster.

In the present analysis we have not accounted for the possible presence of binary stars, that could potentially affect the
estimates of $\overline{\mathrm{RV}}$ and $\sigma_{\mathrm{RV}}$. However, binaries are very rare in GC. According to the study of
\citet{Lucatello15}, who found that the best estimate of the binary fraction among cluster RGB stars is $\sim2\%$, less then one
binary system should statistically be found in the sample of 21 stars with $r<2\arcmin$. In addition, \citet{Bianchini2016} showed
that the incidence of binaries on $\sigma_{\mathrm{RV}}$ is marginal, even when the binary fraction is high, because the velocity
dispersion of the binary system sub-population is reduced by cluster dynamical effects.

Table~\ref{t_RVlit} reports all the measurements of the cluster RV available in the literature. Previous results such as
\citet{Rutledge97} were very likely affected by high field contamination, while more recent estimates converge toward a similar value.
Our result is slightly lower than the latest measurements by \citetalias{Villanova17} and \citet{Gaia18b}.

\begin{table}[t]
\begin{center}
\caption{Literature estimates of the cluster RV.}
\label{t_RVlit}
\begin{tabular}{l l}
\hline
\hline
RV (km~s$^{-1}$) & Reference \\
\hline
\ \ \ $-3\pm9$     & \citet{Mayall46} \\
\ \ \ \ \ $1\pm18$ & \citet{Kinman59} \\
\ \ $1.8\pm6.2$    & \citet{Webbink81} \\
\ \ \ $-2\pm13$    & \citet{Hesser86} \\
$15.9\pm1.2$       & \citet{Pryor89} \\
\ \ \ $32\pm15$    & \citet{Minniti95} \\
\ \ \ $17\pm2$     & \citet{DaCosta95} \\
$42.3\pm10.4$      & \citet{Rutledge97} \\
\ \ \ $13\pm2$     & \citet{Villanova17} \\
$14.6\pm0.8$       & \citet{Gaia18b} \\
$10.5\pm0.5$       & This work \\
\hline
\end{tabular}
\end{center}
\end{table}

\subsection{Cluster membership}
\label{ss_member}

As discussed in Sect.~\ref{ss_rvs}, the spectroscopic sample is substantially contaminated by field stars, which must be removed
to study the cluster metallicity distribution. Unfortunately, the PM information presented in Sect.~\ref{ss_pm} is not a useful
tool in this context. Only 68 targets ($\sim$60\%) fell in the $r<6\arcmin$ field covered by the the PM catalogue. Moreover,
all our targets are saturated in the VVV frames. While the VVV-SkZ\_pipeline can recover an accurate photometry up to
$\approx$3~magnitudes above the saturation level \citep{Mauro13}, the astrometry is more sensitive to saturation, and our VVV PM
catalogue misses 23 ($\sim$34\%) of the brightest remaining objects. In conclusion, we have PM estimates for only 45 targets
($\sim$40\%). In addition, the field-cluster separation in Fig.~\ref{f_pm} is comparable to the error for single stars in most of
the cases. Hence, we did not use the PM information to classify our targets, but we used it {a posteriori} to confirm the
likelihood of our classification.

A first cleaning was performed inspecting the Na~I doublet at 8195~\AA, a feature very sensitive to stellar gravity and prominent
in the spectra of dwarf stars \citep[e.g.][]{Collins16}. The 46 targets showing a very deep and strong doublet were classified as
field dwarfs and excluded from further analysis. They are flagged as `dwarf' in Table~\ref{t_targ}.

Figure~\ref{f_histoRV} shows that a combined selection on radial distance and RV can remove most of the residual contamination by
field giants. To this end, we adopted a two-level selection, similar to that adopted by \citet{Salgado13}, where strict criteria
are used to define the best cluster candidates, and looser criteria for an additional list of probable members. Considering that the
cluster RV dispersion is 7.5~km~s$^{-1}$ (Sect.~\ref{ss_rvs}), and that at $r\approx7\arcmin$ the cluster density is already very
close to the field level (Fig.~\ref{f_radprof}), we first excluded from the analysis the 18 targets with either $r>7\arcmin$ or
$\vert\Delta\mathrm{RV}\vert>16$~km~s$^{-1}$, where $\vert\Delta\mathrm{RV}\vert$ is the difference between the target and
cluster RVs. A flag was assigned to the remaining stars when $r>4\arcmin$, and when $\vert\Delta\mathrm{RV}\vert>8$~km~s$^{-1}$.
Then, the 32 stars with no flag ($r<4\arcmin$ and $\vert\Delta\mathrm{RV}\vert<8$~km~s$^{-1}$) were selected as cluster members,
those with two flags were excluded, and the 17 targets with one flag were considered as an additional set of lower-probability
members. This selection criterion based solely on RVs and radial distance would have excluded all but three stars previously
flagged as foreground dwarfs. Even these three surviving dwarfs (no. 49, 52, and 67) would have been classified only as possible
members of lower probability.

\begin{figure}
\begin{center}
\includegraphics[width=9.cm]{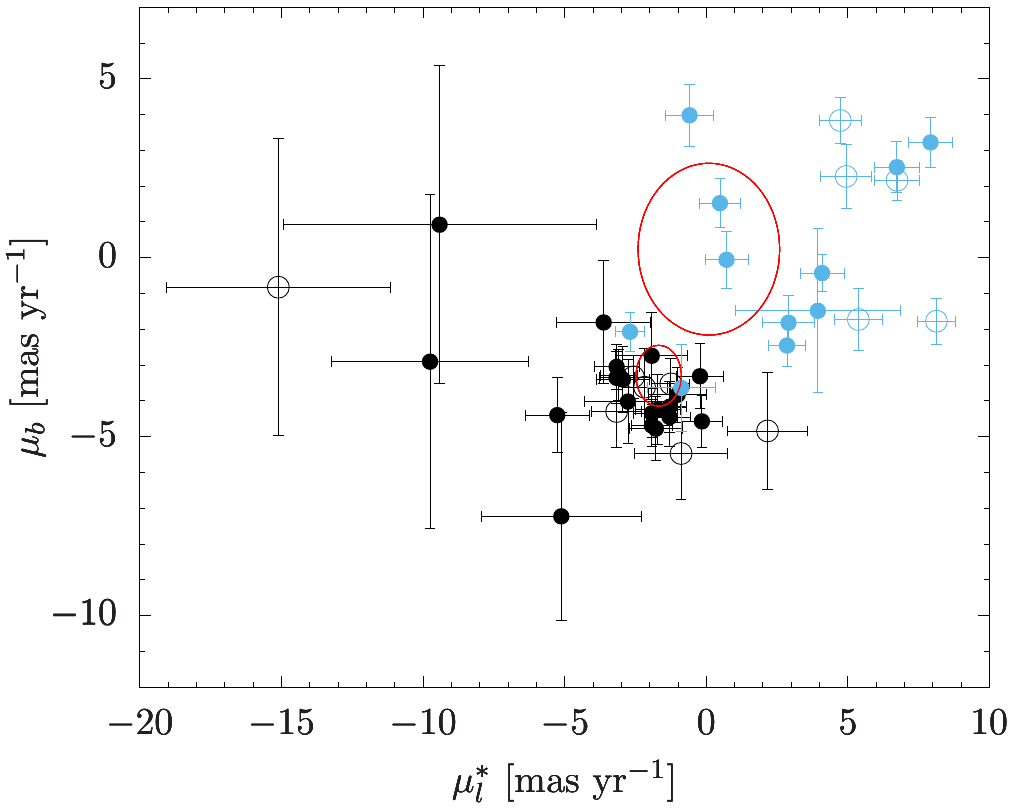}
\caption{Relative proper motion of the spectroscopic targets, according to our classification: dwarf stars and non-members (filled
and empty blue dots), cluster members and probable cluster members (full and empty black dots). The large and small ellipses
are centred on the cluster and field mean value, respectively, with semi-axis equal to the standard deviation of the distributions.}
\label{f_pmclass}
\end{center}
\end{figure}

In Fig.~\ref{f_pmclass} we plot the VVV PM of the 45 targets for which this information is available, indicated with different
symbols according to our classification. The PMs of all stars defined as cluster members or probable members are compatible with
the cluster value, except for a group of targets at large negative $\mu_l^*$ affected by large uncertainties. One object is found
far from the cluster PM, and inside the 1$\sigma$ ellipse of the field PM distribution. This star was removed from the CaT
metallicity analysis. Foreground dwarfs and non-member red giants, on the contrary, are found prevalently at high positive
$\mu_l^*$ far from the cluster value, and only two of them are potentially compatible with cluster membership. Hence, the PMs
confirm that the results of our classification strategy are reliable. We also checked this classification with GAIA parallaxes.
Among the stars classified as members or lower-probability members, we found only two targets whose parallaxes are incompatible
with a distance of 5~kpc at $\sim3\sigma$ level, while 30\% of the objects classified as non-members do not survive this
criterion. The two aforementioned stars were excluded from analysis.

\subsection{CaT measurements}
\label{ss_CaT}

After RV measurements the spectra were shifted to laboratory wavelengths, and then we measured the equivalent widths (EWs) of the
three CaT lines. The CaT feature is a powerful tool for estimating the metallicity of stars in clusters unreachable by
high-resolution spectroscopy. However, as we show later, systematics can arise from differences in the procedure (e.g. continuum
definition and normalisation, EW measurements). It would therefore not be safe to estimate the cluster metallicity from our
measurements without a set of standard objects to correct for the systematics and to adjust the relation between our CaT indexes
and the metallicity scale. Fortunately, the metallicity of M\,28 metallicity has recently been measured from high-resolution
spectra with accuracy \citepalias{Villanova17}. Hence, here we focus only on star-to-star differences, which are independent
from such calibration.

The spectral windows for the continuum definition on both sides of each line were taken from
\citetalias{Rutledge97}. The EWs were measured fitting each observed line with a Voigt profile in a broad band, namely
8490--8506~\AA, 8532--8552~\AA, and 8653--8671~\AA. A variety of algorithms can be found in the literature to combine the three
measurements into one value per star. Some authors prefer the direct sum of all the lines
\citep[e.g.][]{Armandroff88,Olszewski91}, while others use only the two strongest features \citep[e.g.][]{Armandroff91,Saviane12},
or a sum weighted with some specific values \citepalias{Rutledge97}. We experimented with these three alternatives, but identical
results in terms of internal metallicity distribution were obtained in all the cases. Hence, hereafter we adopt the sum of
the three lines ($\Sigma_\mathrm{EW})$ as the final value for each star, but this choice is irrelevant for our conclusions. The
results are given in the eighth column of Table~\ref{t_targ}. No measurement is given for the stars classified as dwarfs in
Sect.~\ref{ss_member}. The EW of these objects was indeed largely discrepant, most of them returning a value more than 40\%
higher than the cluster average.

Our sample comprises four stars in common with \citetalias{Rutledge97}. When we combine the EWs of the three Ca lines with the same
weighted sum adopted in their work, and we compare the results, we find that the slope of our measurements to the R97 measurements
is 0.96. However, our measurements are systematically higher, by 15\% on average. The bands of line and continuum definition are
the same in the two works, but \citetalias{Rutledge97} derived EWs by integration of the observed line profile instead of a fit,
and this is likely the source of the difference. This is in line with the analysis of \citetalias{Rutledge97}, who pointed out that
there is a zero-point offset between the various methods to derive EWs, but the slopes are consistent with unity. Unfortunately,
they do not directly analyse the case of a Voigt profile. Hence, as a consistency check, we repeated the measurements for these
four stars with the same algorithm, but adopting a Gaussian function for the line fit. The resulting values were then corrected by
means of the transformation formula proposed by \citetalias{Rutledge97}. These measurements closely match those of
\citetalias{Rutledge97}, consistent within the errors with unit slope and zero offset.

\begin{figure}
\begin{center}
\includegraphics[width=9.cm]{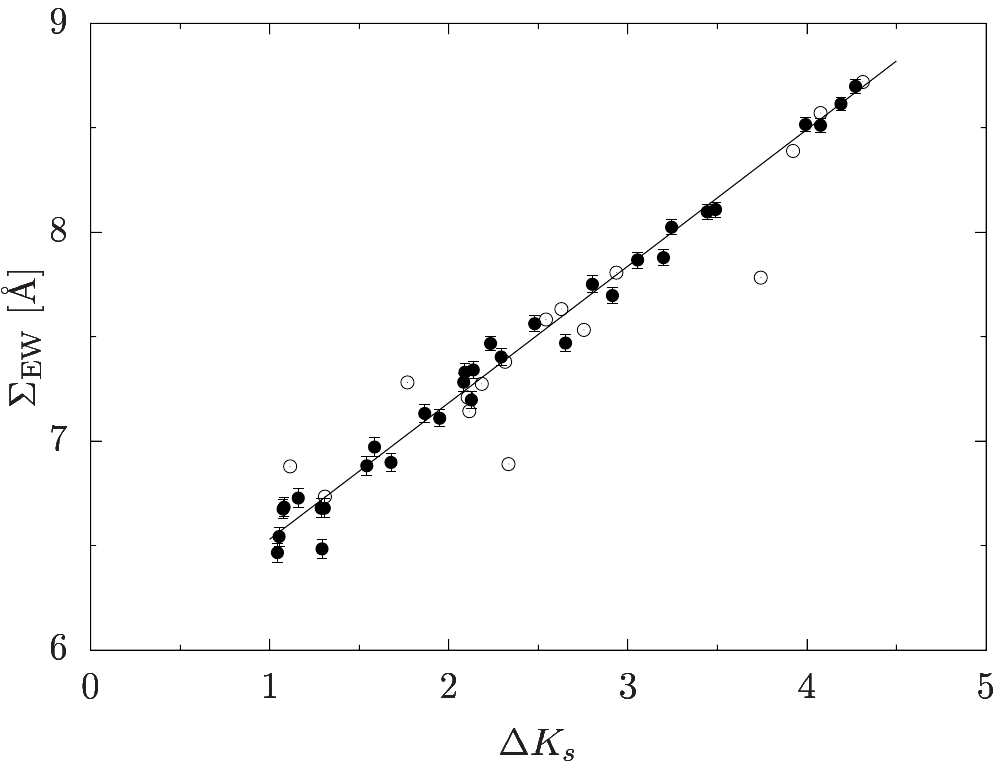}
\caption{Sum of the EW of the three CaT lines ($\Sigma_\mathrm{EW}$) of target stars, as a function of their magnitude difference with
respect to the HB level ($\Delta K_\mathrm{s}$). Full and empty dots are used for the cluster members and the additional possible
cluster members, respectively. The line shows the linear fit of the cluster members.}
\label{f_EWK}
\end{center}
\end{figure}

The strength of the CaT lines depends on the evolutionary stage of the giant star as well as its metallicity. The value of
$\Sigma_\mathrm{EW}$ increases along the RGB, with a linear dependence on magnitude. This degeneracy must be removed to derive an
estimate of metallicity for each individual star. To this end we followed the procedure of \citetalias{Mauro14}. We linearly fit the
position of the 32 cluster stars in the $\Sigma_\mathrm{EW}-\Delta K_\mathrm{s}$ plane, where
$\Delta K_\mathrm{s}=(K_{s,\mathrm{HB}}-K_\mathrm{s})$, and $K_{s,\mathrm{HB}}$ is the magnitude of RGB at the HB level. We adopted
$K_{s,\mathrm{HB}}$=13.0 from \citetalias{Mauro14}. We note that the intersection of HB and RGB is not visible in the cluster CMD
\citep{Cohen17a}, but the exact value of $K_{s,\mathrm{HB}}$ is not relevant as long as we analyse the internal metallicity
distribution. An incorrect definition of this parameter would only result in an identical zero-point offset for all the targets.
We thus found the following solution (see Fig.~\ref{f_EWK}):
\begin{equation}
\Sigma_\mathrm{EW} = (5.903\pm0.031) + (0.646\pm0.012) \times \Delta K_\mathrm{s}.
\label{e_fitCaT}
\end{equation}
The slope is much steeper than that found by \citetalias{Mauro14} combining the same VVV photometric
data with spectroscopic measurements of \citet[][0.385~\AA~mag$^{-1}$]{Saviane12}, and \citetalias{Rutledge97} (0.348~\AA~mag$^{-1}$).
Part of this difference is due to the adopted algorithm for $\Sigma_\mathrm{EW}$ because our direct sum of three EWs increases more
with line strength then the sum of only two features \citep{Saviane12}, or a weighted mean that reduces the contribution of some of
them \citepalias{Rutledge97}. We verified that the slope reduces to $\approx$0.53~\AA~mag$^{-1}$ if we adopt these algorithms for
$\Sigma_\mathrm{EW}$. Another factor affecting the slope of the relation is the method used to measure EWs. When fitting
the lines with a Gaussian profile, and correcting the results with the aforementioned equations given by \citetalias{Rutledge97},
we found a slope of 0.387~\AA~mag$^{-1}$, very similar to the result of \citetalias{Mauro14}.

The slope of Eq.~(\ref{e_fitCaT}) indicates how $\Sigma_\mathrm{EW}$ grows with $\Delta K_\mathrm{s}$ at a fixed metallicity,
while for a fixed $\Delta K_\mathrm{s}$ the value $\Sigma_\mathrm{EW}$ depends on metallicity alone. To study the internal
metallicity distribution of cluster stars, we therefore derived for each star the vertical distance from the cluster relation
(Equation~\ref{e_fitCaT}) in the $\Sigma_\mathrm{EW}-\Delta K_\mathrm{s}$ plane:
\begin{equation}
\Delta\Sigma_\mathrm{EW}=\Sigma_\mathrm{EW}-\Sigma_\mathrm{EW,fit}(\Delta K_\mathrm{s}).
\label{e_DeltaS}
\end{equation}

We note that the star 66 is an outlier among cluster members at the fainter end of the sample, with
$\Delta K_\mathrm{s}\approx$1.4 and $\Delta\Sigma_\mathrm{EW}<-250$~m\AA\footnote{For a more readable notation, in this section and
hereafter we use the units m\AA, where 1~m\AA=$10^{-3}$\AA.}. This target was removed from the fit, and transferred to the group of
possible additional cluster members in the following analysis. However, if this target is included, the zero-point and slope of
Equation~\ref{e_fitCaT} change by only 1$\sigma$ ($\Sigma_\mathrm{EW}=5.872+0.656\times\Delta K_\mathrm{s}$).

The main source of uncertainty on $\Delta\Sigma_\mathrm{EW}$ is the measurement error on the EWs of the three lines. To estimate it,
we summed the spectra of the eight brightest targets with $K_\mathrm{s}<9.1$ (excluding those flagged as non-members or foreground
dwarfs) to create a high-quality template spectrum. This was then multiplied by a linear function of random slope to alter its
continuum, and degraded to S/N=90, 170, and 250, to match the spectral quality of the observed spectra of the faintest, intermediate,
and brightest targets, respectively. The same measurement routine was run on these artificial spectra as for the real data, and
the procedure was repeated 100~times. The results showed a rms scatter decreasing with S/N, namely 45, 40, and 32~m\AA\ for the three
cases, that was assumed as the measurement error on $\Delta\Sigma_\mathrm{EW}$. We also repeated the whole procedure adopting as a
template a synthetic spectrum drawn from the \citet{Coelho05} library, with $T_\mathrm{eff}$=4250~K, $\log{g}$=1.5,
$[\frac{\mathrm{Fe}}{\mathrm{H}}]=-1.5$, and $[\frac{\alpha}{\mathrm{Fe}}]=+0.4$, and we found very similar results.

Another source of error on $\Delta\Sigma_\mathrm{EW}$ are the uncertainties on the fit parameters in Equation~(\ref{e_fitCaT}). We
estimated these measuring how the dispersion of $\Delta\Sigma_\mathrm{EW}$ varied when the parameters were altered by $\pm 1\sigma$.
We found that it quadratically increased by 9~m\AA\ at all magnitude ranges. The photometric uncertainties on $K_\mathrm{s}$ are in
the range 0.010--0.015~magnitudes, and they propagate into an error on $\Delta\Sigma_\mathrm{EW}$ smoothly increasing with magnitude
from 6 to 10~m\AA. Quadratically summing the measurement errors, and the uncertainties introduced by the fitting procedure and by
photometric errors, we eventually obtained the final errors $\sigma_{\Delta\Sigma}$=34, 42, and 47~m\AA\ in the bright, intermediate,
and faint magnitude ranges, respectively.

\subsection{Internal metallicity distribution}
\label{ss_metspread}

\begin{figure}
\begin{center}
\includegraphics[width=7.cm,angle=-90]{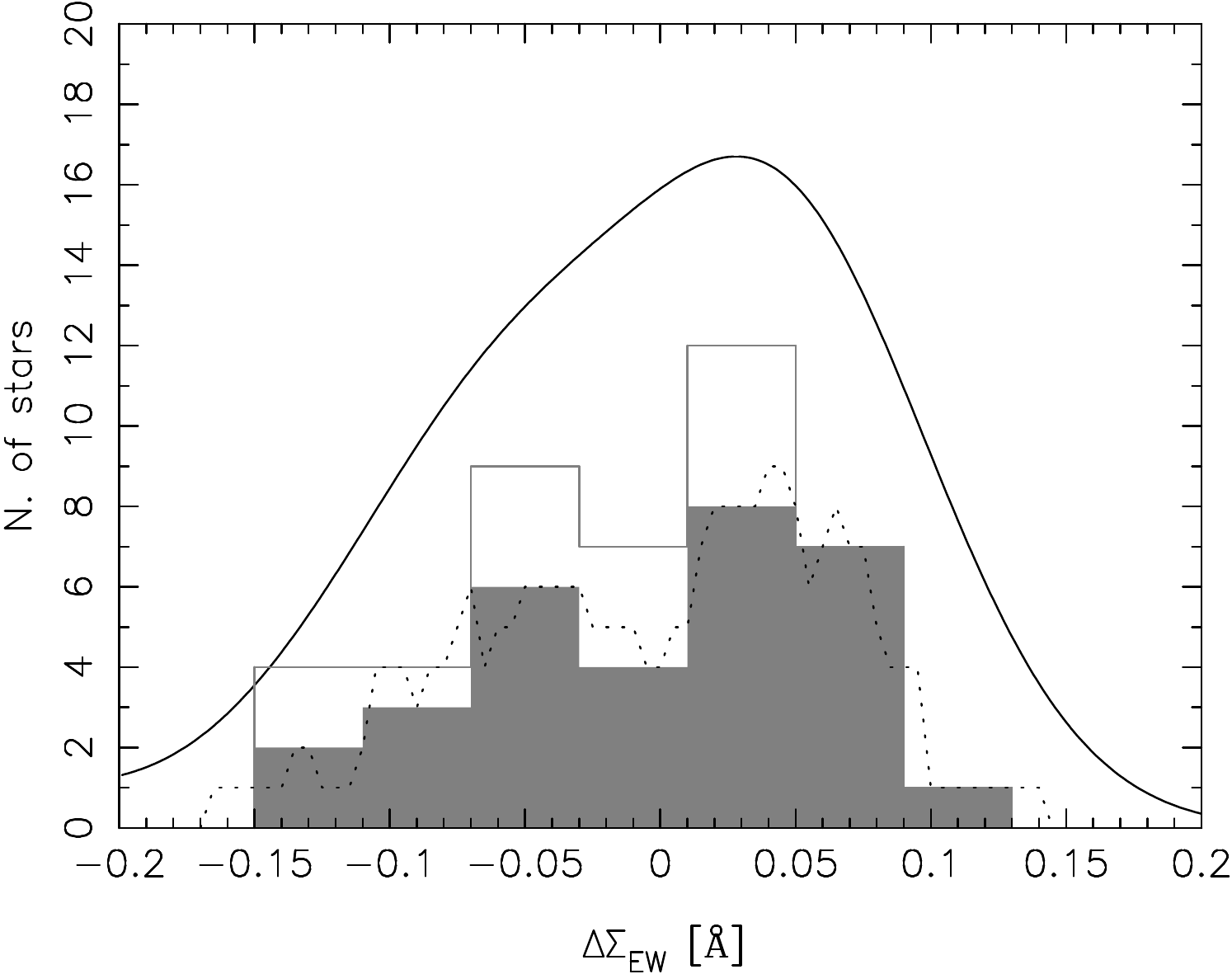}
\caption{Histogram of distribution of $\Delta\Sigma_\mathrm{EW}$ for the observed targets. Different algorithms are shown, as
detailed in the text.}
\label{f_histoDSEW}
\end{center}
\end{figure}

The global observed dispersion of $\Delta\Sigma_\mathrm{EW}$ is $\sigma_{\Delta\Sigma,\mathrm{obs}}$=66~m\AA, comparable but
slightly larger than the estimated errors $\sigma_{\Delta\Sigma}$. However, $\sigma_{\Delta\Sigma,\mathrm{obs}}$=11~m\AA\ for the
five brightest stars with $\Delta K_\mathrm{s}>4$, much smaller than $\sigma_{\Delta\Sigma}$, while it slightly increases at fainter
magnitudes from 71~m\AA\ for $\Delta K_\mathrm{s}=$2--4 to 74~m\AA\ for the ten faintest objects. If the measurements are scattered
by an intrinsic spread, this should be of the order of $\sqrt{\sigma_{\Delta\Sigma,\mathrm{obs}}^2-\sigma_{\Delta\Sigma}^2}$=57~m\AA\
at all magnitudes fainter than $\Delta K_\mathrm{s}=2$.

The distribution histogram of $\Delta\Sigma_\mathrm{EW}$ is shown in Fig.~\ref{f_histoDSEW}. The dark shaded area indicates a
classical histogram in bins of 40~m\AA\ for the 32 stars considered cluster members, and the empty histogram is obtained including
the 17 additional member candidates. The dotted line is the result of overlapping bins of the same width, in steps of 10~mA. The
solid curve shows a continuous function obtained substituting to each star a Gaussian centred on the corresponding
$\Delta\Sigma_\mathrm{EW}$, with width $\sigma$ equal to the mean observational error (45~m\AA), and unit maximum height. One star
classified as a cluster member and four lower-probability members are not shown in the figure because they are found at
$\vline\Delta\Sigma_\mathrm{EW}\vline>200$~m\AA, and they will not be considered further.

The histograms of Fig.~\ref{f_histoDSEW} suggest a bimodal distribution, with two peaks at approximately
$\Delta\Sigma_\mathrm{EW}=\pm 50$~m\AA, symmetric with respect to the mean cluster value ($\Delta\Sigma_\mathrm{EW}=0$).
Nevertheless, a series of independent tests agreed that this bimodality has no statistical significance. The Shapiro-Wink test
\citep{Shapiro65} is considered one of the most powerful tools in the literature to test the normality of a distribution of small
samples. The related Shapiro-Wink statistics for our 31 cluster members is $W=0.967$, and it indicates that the data have a
$\sim$47\% probability of being normally distributed. This decreases to $\sim$33\% if the 13 additional stars are included.
The KMM algorithm\footnote{https://www.pa.msu.edu/ftp/pub/zepf/kmm/} \citep{McLachlan87,Ashman94} is somehow more optimistic,
assigning the null hypothesis of a single Gaussian distribution a probability of 13\% (20\% when the 13 additional
candidate members are included). KMM is a maximum-likelihood algorithm that assesses the improvement in fitting a multi-group
model over a single-group model, and it has been repeatedly applied to astrophysical datasets
\citep[e.g.][]{Ostrov93,Lee93,Bird94}. A Kolmogorov-Smirnov test indicates that the $\Sigma_\mathrm{EW}$ of the 31 members has
a 55\% probability of being drawn from a Gaussian distribution with $\sigma=66$~m\AA, which increases to 75\% if the additional
member candidates are also considered. This suggests no significant statistical evidence of bimodality. In addition, if we compare
the distributions with a Gaussian of width $\sigma$=45~m\AA\ (equalling the estimated errors), the same probability is reduced but
still acceptable (18\% for the member stars, and 13\% when the additional candidates are included). Hence, the test negates a
statistical significance even to the observed dispersion $\Delta\Sigma_\mathrm{EW}$ being wider than the estimated errors.

It would be very instructive to translate the values of $\Delta\Sigma_\mathrm{EW}$ into [$\frac{\mathrm{Fe}}{\mathrm{H}}$], to
estimate the errors, the possible internal spread, and the separation between the two peaks of a bimodal distribution, directly
in a metallicity scale. However, we could not use $\Delta\Sigma_\mathrm{EW}$--[$\frac{\mathrm{Fe}}{\mathrm{H}}$] calibrations
from the literature \citepalias[e.g.][]{Mauro14} because our EW measurements are not on the same scale of the previous works,
as shown in Sect.~\ref{ss_CaT}. We obtained a rough metallicity scale for our measurements from the stars in common with
\citetalias{Rutledge97}. We first derived a metallicity value for each of them by means of the \citetalias{Mauro14} calibration;
then we used them to fit a linear relation between [$\frac{\mathrm{Fe}}{\mathrm{H}}$] and our $\Delta\Sigma_\mathrm{EW}$ values.
The fit indicates that $\Delta\Sigma_\mathrm{EW}=100$~m\AA\ corresponds to about
$\Delta$[$\frac{\mathrm{Fe}}{\mathrm{H}}$]$\approx0.15$~dex. This rough result is reasonable, as it indicates that our estimate
of the observational errors ($\sim$45~m\AA) equals $\approx$0.07~dex in metallicity, which is closely comparable with the
accuracy expected by \citetalias{Mauro14} for relative CaT measurements. According to this scale the two peaks observed in
Fig.~\ref{f_histoDSEW} would be separated by $\approx$0.13--0.14~dex, and the spread of 57~m\AA\ enhancing the observed
distribution compared to the estimated errors would be about 0.08~dex.


\section{Discussion}
\label{s_disc}

\subsection{Dynamical mass}
\label{ss_mass}

\begin{figure}
\begin{center}
\includegraphics[width=9.cm]{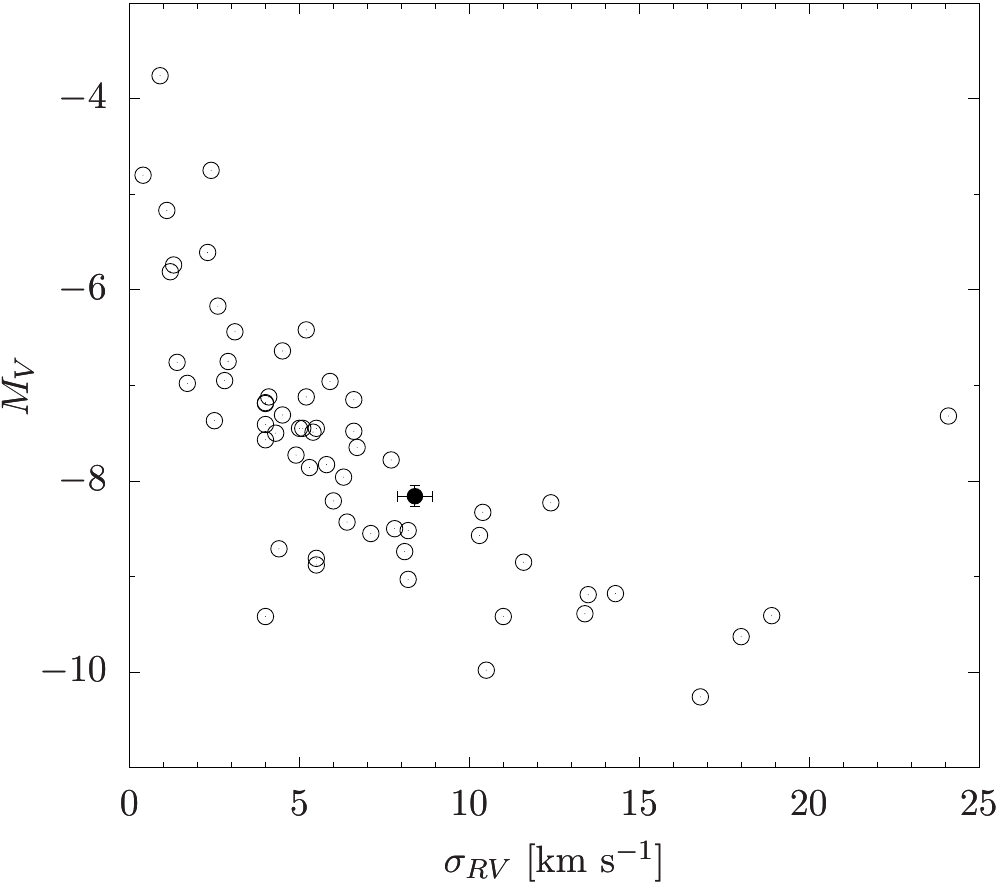}
\caption{Absolute luminosity in the $V$ band as a function of the central RV dispersion of all GCs in the \citet{Harris96}
catalogue with available measurements. The black dot represents M\,28, with our dispersion measurement.}
\label{f_rvdispMv}
\end{center}
\end{figure}

The large velocity dispersion found in Sect.~\ref{ss_rvs} is in line with the high cluster total luminosity.
In Fig.~\ref{f_rvdispMv}, we show the central RV dispersion and the absolute integrated magnitude in the $V$ band for all the
Galactic GCs for which these estimates are available. All data were drawn from the \citet[][]{Harris96} catalogue, except for
M\,28, where we compare our RV dispersion with the average of the available absolute magnitude estimates
\citep{Webbink85,Peterson87,vandenberg91}. The plot shows that our result closely fits the $\sigma_\mathrm{RV}-M_\mathrm{V}$
relation for Galactic GCs. However, we note that our estimate should still be regarded as a lower limit for the central value
because it is obtained for stars distributed in the inner $2\arcmin$ from the centre.

The large velocity dispersion suggests that M\,28 should be a massive cluster. To confirm this we estimated its dynamical mass
from the \citet{King66} model, as modified by \citet[][Eq.~9]{Illingworth76}, assuming $r_t=13\farcm0\pm1\farcm4$,
$d=5.01\pm0.23$~kpc, and $\sigma_{\mathrm{RV,c}}=8.4\pm0.5$~km~s$^{-1}$ from this work, and $r_c=0\farcm24\pm0\farcm2$ from the
literature (see Sect.~\ref{ss_raddens}). The model parameter $\mu$ was calculated from the tidal and core radii as in
\cite{Illingworth76b}. We thus find $\log{\frac{\mathrm{M}}{\mathrm{M}_\odot}}=5.15\pm0.12$. Alternatively, we find
$\log{\frac{\mathrm{M}}{\mathrm{M}_\odot}}=5.51\pm0.09$ from the \citet{deVaucouleurs59} model \citep[][Eq.~13]{Illingworth76},
with the half-light radius $r_e=2.25\pm0.37$~pc \citep{vandenberg91}.

\citet{Mandushev91} showed that the simplified single-mass King model at the basis of our estimate should underestimate
$\log{\frac{\mathrm{M}}{\mathrm{M}_\odot}}$ by about 0.3~dex with respect to more elaborated multi-component King models. It
should be noted that if we applied this offset to our result, our two independent estimates would actually coincide. Very
interestingly, our value based on this single-mass King model is very similar to the result obtained by \citet{Mandushev91}
with the same method, while the higher value returned by the de Vaucouleurs model matches the more recent estimate of
\citet{Baumgardt18} based on more elaborated N-body simulations.

We do not correct our result obtained from the King model for the expected underestimate, but we do assume the weighted average
of our two results as our best estimate for the cluster mass, $\log{\frac{\mathrm{M}}{\mathrm{M}_\odot}}=5.38\pm0.15$, which is
intermediate between the results of \citet{Mandushev91} and \citet{Baumgardt18}. According to these two compilations, this value
classifies M\,28 as one of the 30\% most massive systems among the 147 catalogued Galactic GCs.

\subsection{Bulge age-metallicity relation}
\label{ss_amr}

\begin{figure}
\begin{center}
\includegraphics[width=9.cm]{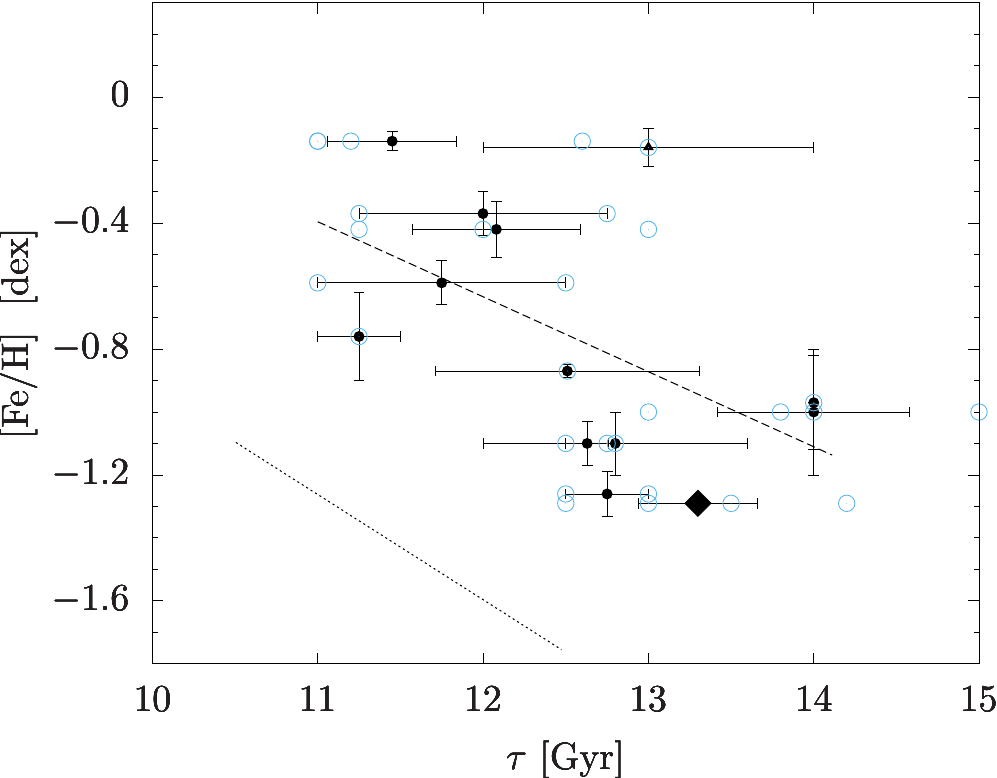}
\caption{Age-metallicity relation of bulge clusters. The dashed line indicates the linear fit of the data, while the dotted
line shows a similar relation for halo objects. Empty light blue dots are used for the single literature measurements, while
full black dots show the average value for each cluster. The outlier NGC\,6553 is shown as a triangle, while M\,28 is shown
as a larger diamond.}
\label{f_agemet}
\end{center}
\end{figure}

Our analysis of VVV photometry indicates that M\,28 is a very old cluster, indeed one of the oldest objects in the Galactic
bulge, as previously suggested in the literature \citep{Testa01,Roediger14,Villanova17,Kerber18}. On the other hand, while some
metal-poor halo intruders can be found in the inner Galactic regions \citep[e.g.][]{Myeong19}, M\,28 is also one of the most
metal-poor objects catalogued as a genuine bulge GC by \citet{Bica16}. To put this result into context, in Fig.~\ref{f_agemet}
we compare the age and metallicity estimates available in the literature for bulge GCs. We included NGC\,6569 as a genuine
bulge cluster, although \citet{Bica16} classify it as an `outer shell' object, while we excluded Terzan\,5, because its
complex stellar population suggests that its origin is probably distinct from that of the other objects in the sample
\citep{Lanzoni10}. For metallicity, we based our value on the most recent high-resolution spectroscopic measurements whenever
possible, namely for HP1 \citep{Barbuy16}, M\,28 \citepalias{Villanova17}, NGC\,6522 \citep{Barbuy09},
NGC\,6528 \citep{Munoz2018}, NGC\,6558 \citep{Barbuy07}, and NGC\,6569 \citep{Johnson18}. For all other objects
(NGC\,6304, NGC\,6553, NGC\,6624, NGC\,6637, NGC\,6652, NGC\,6717, NGC\,6723), we relied on the compilation of
\citet{Carretta09b}. The main reference for age estimates were
\citet[][NGC\,6304, NGC\,6624, NGC\,6637, NGC\,6717, and NGC\,6723]{Dotter10} and
\citet[][NGC\,6304, NGC\,6624, NGC\,6637, NGC\,6652, NGC\,6717, NGC\,6723]{VandenBergh13}, plus additional measurements for
single clusters, namely HP1 \citep{Kerber19}, NGC\,6522 \citep{Meissner06,Barbuy09,Kerber18}, NGC\,6528
\citep{Calamida14,Felzing02,Momany03,Lagioia14}, NGC\,6553 \citep{Zoccali01}, NGC\,6558 \citep{Barbuy07}, NGC\,6569
\citep{Saracino19}, and NGC\,6624 \citep{Saracino16}. When more than one measurement is available, in Fig.~\ref{f_agemet} we
show the mean value, with the error-on-the-mean as an associated uncertainty.

Figure~\ref{f_agemet} shows that an age-metallicity relation (AMR) is most likely present among bulge GCs: all objects with
[Fe/H]$\la-$0.8 are older than $\sim$12.5~Gyr, while all but one of the more metal-rich GCs are younger than $\sim$12.0~Gyr.
The only exception to this general trend is NGC\,6553, which is one of the oldest (13$\pm$1~Gyr) and most metal-rich
([Fe/H]=$-0.16\pm0.06$) clusters in the sample. The AMR is clear but not well defined, because the objects span a rather
narrow range in age, and the age measurements are affected by large errors. These uncertainties also blur the intrinsic
dispersion that might indicate the degree of inhomogeneity in the chemical composition at early epochs. While metallicities are
in general accurately measured by high-resolution spectroscopic studies, the use of different model isochrones introduces systematics
in the age estimates \citep[see e.g.][]{Kerber18}, and a homogeneous determination for all the objects would be required.
Even so, we observe that as the bulk of the age determinations come from a handful of works with many objects in common, the
systematics between them tend to cancel out. An AMR is observable in Fig.~\ref{f_agemet} even if it considers only the
single measurements instead of averages, although with an increased dispersion.

In Fig.~\ref{f_agemet} the AMR of the Galactic halo GCs is also shown for comparison. This was obtained with a linear fit
of age and metallicities from the \citet{VandenBergh13} compilation, considering objects classified as halo members by
\citet{Leaman13}, and excluding members of the Sagittarius galaxy \citep[e.g.][]{Mottini08}. The comparison shows that the
bulge GC system clearly formed and evolved extremely fast. At the earliest stages, 13-14~Gyr ago, it formed clusters 5-10 times
more metal rich than in the halo; then it further increased its metal content by one order of magnitude in about 2 Gyr.

\subsection{Cluster orbit}
\label{ss_orbit}

\begin{figure}[]
\begin{center}
\includegraphics[width=9.cm]{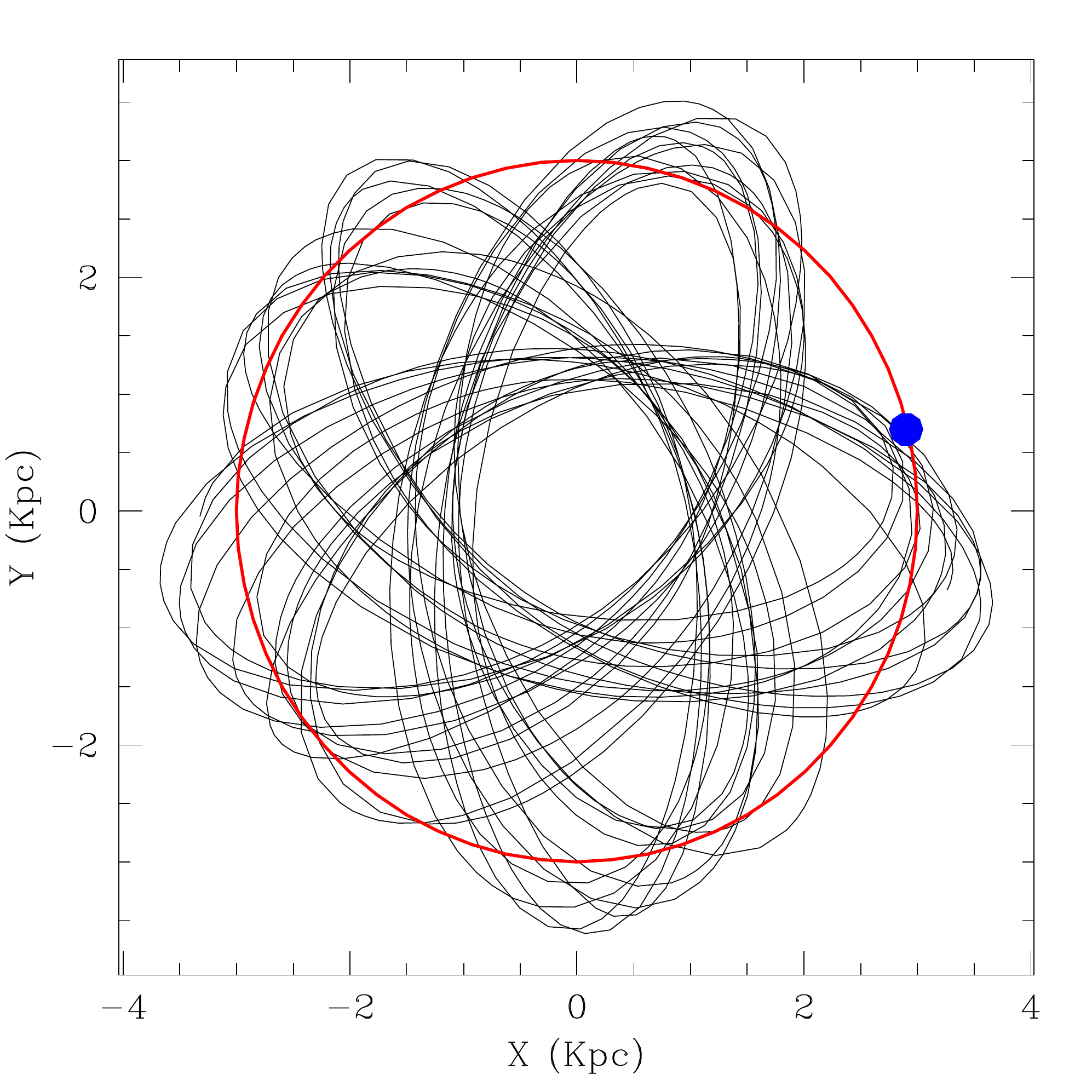}
\includegraphics[width=9.cm]{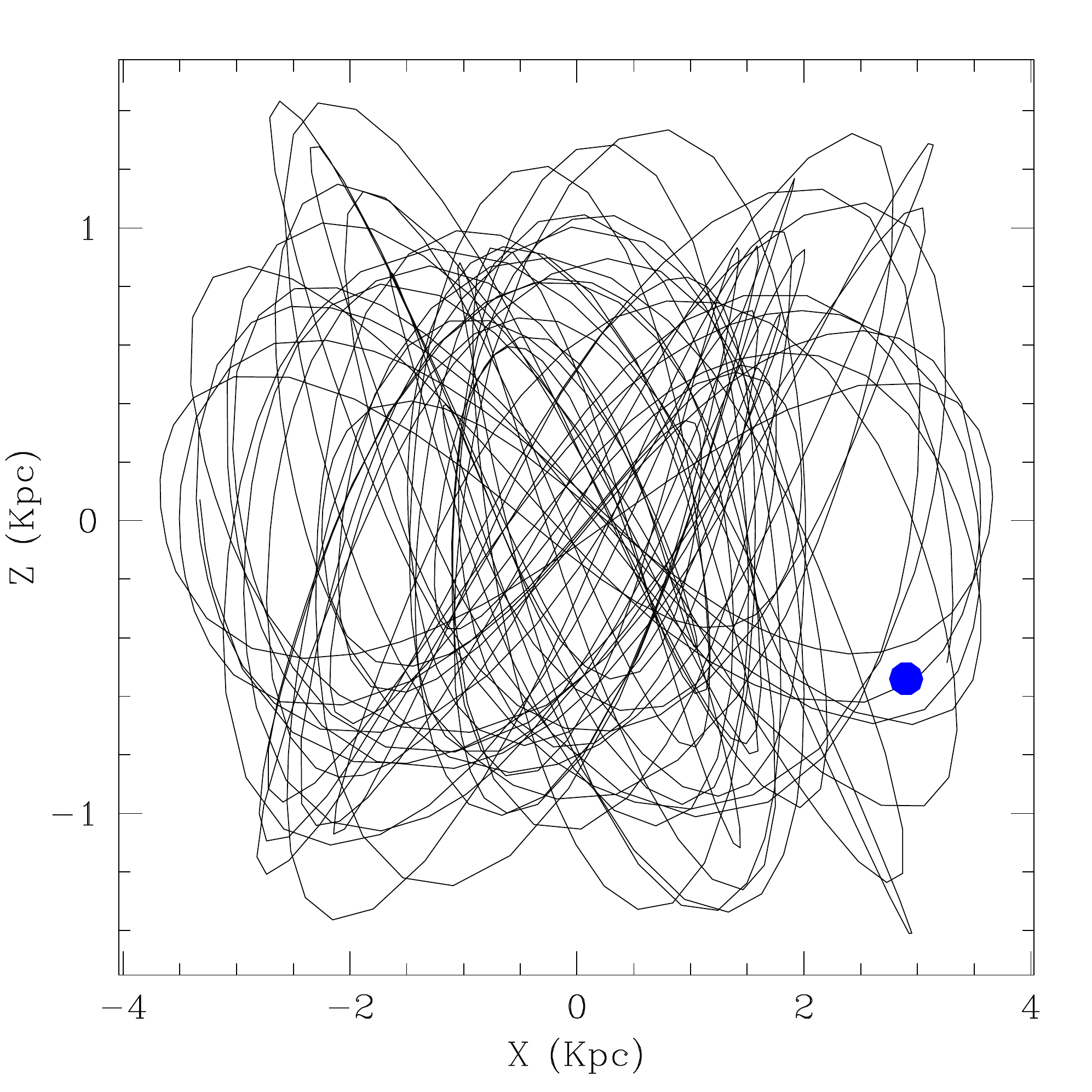}
\caption{Orbit of M\,28 in Galactic cartesian coordinates, in the $XY$ ({\it upper panel}) and $XZ$ ({\it lower panel}) plane.
The blue dot gives the actual position of the cluster, and the red circle indicates the distance of 3~kpc from the Galactic
centre.}
\label{f_orbit}
\end{center}
\end{figure}

The first orbit calculation for M\,28 was presented by \citet{Dinescu99}, who found a thick-disk solution. However, this result
was based on the early PM measurements of \citet{Cudworth93}, which are not confirmed by more recent estimates
(see Table~\ref{t_pm}). \citet{Dinescu13} later presented a new solution based on updated measurements, and found an eccentric
and disruptive orbit, confined within $\sim$3.0~kpc of the Galactic centre, with a pericentre at only R$\sim$0.5~kpc. The
authors also suggested that M\,28 is currently close to the apocentre. \citet{Chemel18} started from quite different PMs (see
Table~\ref{t_pm}), but obtained similar results, although their solution is less eccentric.

We calculated again the orbit solution for this cluster in light of our new results for RV, PM, and distance presented in
Table~\ref{t_M28res}. This task was performed with the web-code {\it GravPot16}\footnote{https://gravpot.utinam.cnrs.fr/}
\citep{Fernandez17}, whose calculation is based on a Galactic gravitational potential driven by the Besançon Galaxy Model mass
distribution \citep{Robin03}. GravPot16 includes a prolate bar, as modelled in \citet{Robin12}, and this is important for all
those clusters like M28 whose orbit enters into or is confined to this region of the Galaxy. We assumed R$_{\odot}$= 8.2~kpc and
(U$_{\odot},V{\odot},W{\odot}$)=(11,252,7)~km~s$^{-1}$ as solar radius and solar velocity components \citep{Schonrich10}. We
integrated the orbit forward for 3~Gyr. We report the projection of the orbit on the (X,Y) and (X,Z) Galactic planes in
Fig.~\ref{f_orbit}. The upper panel shows that the orbit is confined within the bulge region. The cluster minimum distance from
the Galactic centre is 0.8~kpc, while the maximum distance is 3.8~kpc. The orbit in the lower panel, on the other hand, suggests
a maximum height of about 1.4~kpc. The eccentricity and the Z-component of the orbital angular momentum (L$_Z$) are 0.49 and
0.3~kpc~km~s$^{-1}$, respectively.


\section{Conclusions}
\label{s_concl}

\begin{table}[t]
\begin{center}
\caption{Derived cluster parameters.}
\label{t_M28res}
\begin{tabular}{l l}
\hline
\hline
($\alpha$,$\delta$) & (18:24:32.58, $-$24:52:13.6) \\
$r_t$ & $13\farcm0\pm1\farcm4$ \\
$\overline{\mathrm{RV}}$ & 10.5$\pm$0.5~km~s$^{-1}$ \\
$\sigma_{\mathrm{RV,c}}$ & 8.4$\pm$0.5~km~s$^{-1}$ \\
$E(J-K_\mathrm{s})$ & $0.20\pm0.02$ \\
$(m-M)_{K_\mathrm{s}}$ & $13.6\pm0.1$ \\
$d$ & $5.01\pm0.23$~kpc \\
$\tau$ & 13--14~Gyr \\
$(\mu_\alpha^*,\mu_\delta)$ & $(-0.35\pm0.08,-8.54\pm0.08)$~mas~yr$^{-1}$ \\
$\log{\frac{\mathrm{M}}{\mathrm{M}_\odot}}$ & $5.38\pm0.15$ \\
\hline
\end{tabular}
\end{center}
\end{table}

We have analysed wide-field VVV photometric data, VVV proper motions, and intermediate-resolution CaT spectroscopy to study
the properties of the bulge globular cluster M\,28. The main cluster parameters obtained in this work are presented in
Table~\ref{t_M28res}. Our results in general agree with previous measurements in the literature, confirming the properties
of this cluster in terms of reddening, distance, and angular size. We also find that the recent estimates of the cluster
RV are converging to a value in the range $\overline{\mathrm{RV}}=11-14$~km~s$^{-1}$, slightly lower than the
17~km~s$^{-1}$ quoted in the \citet[][December 2010 web version]{Harris96} catalogue. The orbit, calculated from our
measurements of RV, PM, and distance, is confined within 3.8~kpc of the Galactic centre, and 1.4~kpc of the plane. M\,28 is
therefore a genuine bulge cluster.

The internal RV dispersion of M\,28 is high, with a central value of $\sigma_{\mathrm{RV,c}}=8.4\pm0.5$~km~s$^{-1}$, from
which we obtained a rough estimate of the total cluster mass of $\log{\frac{\mathrm{M}}{\mathrm{M}_\odot}}=5.38\pm0.15$. This
should still be an underestimate, both because $\sigma_{\mathrm{RV,c}}$ was estimated from stars up to $2\farcm0$ from the
centre, and because of a probable underestimate introduced by the use of a single-mass King model, as discussed in
Sect.~\ref{ss_mass}. This result places M\,28 among the 30\% most massive GCs in the Milky Way.

M\,28 is undergoing mass loss as a consequence of Galactic tidal forces and shock effects. We detected a tail of stars
trailing the cluster in its motion, and a structure directed toward the Galactic plane, as expected by models of GC tidal
tails formations \citep[e.g.][]{Kupper10,Montuori07}. In particular, the structures directed toward the Galactic plane and
centre are a consequence of relevant tidal shocks from the disk and bulge. Then, despite its present-day high mass, M\,28 must
have been an even larger structure in the past.

We find that this cluster, with an age of 13-14~Gyr, is one of the oldest objects in the Galactic bulge. We put this result
in the context of the bulge evolution history, comparing the age and the metallicity of all bulge GCs with reliable
measurements in the literature. The resulting AMR demonstrates that the bulge system evolved quickly, forming clusters 5-10
times more metal-rich than in the halo at the earliest stages, then increasing its metal content by one order of magnitude in
about two Gyrs.

Our CaT measurements show no hint of any metallicity inhomogeneity inside the cluster. We found no statistical significance of
a double distribution of iron content, and we derived an upper limit to any internal metallicity spread of 0.08~dex. We thus
strengthen with a larger statistical sample (32 stars) the result found by \citetalias{Villanova17} in their high-resolution analysis
of 17 red giant stars. Hence, M\,28 can be considered a classical GC, with large star-to-star variations of light elements
\citepalias{Villanova17}, but a homogeneous metallicity.

In conclusion, the age and metallicity of M\,28, its orbit, and the absence of chemical peculiarities \citepalias{Villanova17},
support the \citet{Bica16} classification of a genuine bulge GC. However, its very old age, its high mass, and its evident mass
loss suggest that this cluster should be the remnant of a larger structure, possibly a very massive GC or a primeval bulge
building block, as previously suggested by \citet{Lee07} and \citet{Chun12}.


\begin{acknowledgements}
The authors thanks the referee for their very helpful comments and suggestions.
C.M.B., S.V., and D.M. acknowledge support from FONDECYT through regular projects 1150060, 1170518, and 1170121. D.M. and D.G.
gratefully acknowledge support from the BASAL Centre for Astrophysics and Associated Technologies (CATA) through grant AFB
170002. D.G. also acknowledges financial support from the Direcci\'on de Investigaci\'on y Desarrollo de la Universidad de La
Serena through the Programa de Incentivo a la Investigaci\'on de Acad\'emicos (PIA-DIDULS). Support to J.B. and M.Z. is
provided by the Ministry for the Economy, Development and Tourism, Programa Iniciativa Cient\'ifica Milenio grant IC120009,
awarded to the Millennium Institute of Astrophysics (MAS), and by ANID, Millennium Science Initiative ICN12\_009 awarded to the
Millennium Institute of Astrophysics (MAS). A.-N.C. is supported by the Gemini Observatory, which is operated by the Association
of Universities for Research in Astronomy, Inc., on behalf of the international Gemini partnership of Argentina, Brazil, Canada,
Chile, and the United States of America. We gratefully acknowledge data from the ESO Public Survey program ID 179.B-2002 taken
with the VISTA telescope, and products from the Cambridge Astronomical Survey Unit (CASU). This work has made use of data from
the European Space Agency (ESA) mission {\it Gaia} (https://www.cosmos.esa.int/gaia), processed by the {\it Gaia} Data
Processing and Analysis Consortium (DPAC, https://www.cosmos.esa.int/web/gaia/dpac/consortium). Funding for the DPAC has been
provided by national institutions, in particular the institutions participating in the {\it Gaia} Multilateral Agreement. This
investigation made use of data from the Two Micron All Sky Survey, which is a joint project of the University of Massachusetts
and the Infrared Processing and Analysis Center/California Institute of Technology, funded by the National Aeronautics and Space
Administration and the National Science Foundation.
\end{acknowledgements}


\bibliographystyle{aa}
\bibliography{M28Cat}


\clearpage
\onecolumn

\begin{longtable}{lccrcrrl}
\caption{Data of the observed target stars.
\label{t_targ}} \\
\hline
ID & RA & Dec & $K_\mathrm{s}$ & $(J-K_\mathrm{s})$ & RV & $r$ & $\Sigma_{EW}$ \\
 & \tiny{hh:mm:ss.s} & \tiny{ $^o$: $\arcmin$: $\arcsec$} & & & \tiny{km~s$^{-1}$} & \tiny{$\arcmin$} & \tiny{\AA} \\
\hline
\endfirsthead
\caption{continued.}\\
\hline
ID & RA & Dec & $K_\mathrm{s}$ & $(J-K_\mathrm{s})$ & RV & $r$ & $\Sigma_{EW}$ \\
\hline
\endhead
\hline
\endfoot
4   & 18:24:18.6 & $-$24:43:47 & 11.09 & 0.84 &  $-29.1\pm1.0$ &  9.0 & dwarf \\
5   & 18:24:31.4 & $-$24:51:14 &  8.95 & 1.09 &    $7.9\pm0.6$ &  1.0 & 8.581 \\
6   & 18:24:26.7 & $-$24:50:22 & 11.95 & 0.79 &    $8.5\pm0.6$ &  2.3 & 6.543 \\
8   & 18:24:30.2 & $-$24:51:05 & 11.71 & 0.81 &   $15.6\pm0.7$ &  1.3 & 6.678 \\
9   & 18:24:26.6 & $-$24:51:13 &  9.01 & 1.06 &   $15.8\pm0.6$ &  1.8 & 8.516 \\
10  & 18:24:27.9 & $-$24:51:16 & 10.87 & 0.87 &   $-0.8\pm0.7$ &  1.5 & 7.197 \\
11  & 18:24:17.0 & $-$24:48:25 & 11.69 & 0.82 &  $-52.6\pm0.7$ &  5.3 & 8.274 \\
12  & 18:24:37.0 & $-$24:45:55 & 10.82 & 0.88 &   $11.3\pm0.7$ &  6.3 & 7.274 \\
13  & 18:24:37.2 & $-$24:44:33 & 10.40 & 0.87 &  $-20.0\pm1.1$ &  7.7 & dwarf \\
14  & 18:24:33.0 & $-$24:48:10 &  9.26 & 1.02 &    $1.1\pm0.5$ &  4.0 & 7.783 \\
15  & 18:24:31.8 & $-$24:48:29 &  8.73 & 1.15 &    $5.2\pm0.7$ &  3.7 & 8.698 \\
16  & 18:24:29.5 & $-$24:43:19 & 11.97 & 0.82 &  $-18.6\pm0.9$ &  8.9 & 7.669 \\
17  & 18:24:27.9 & $-$24:44:17 & 11.09 & 0.86 &  $-26.8\pm1.3$ &  8.0 & dwarf \\
18  & 18:24:27.7 & $-$24:40:41 & 10.88 & 0.84 &   $53.6\pm0.8$ & 11.6 & dwarf \\
19  & 18:24:25.6 & $-$24:45:15 & 10.87 & 0.87 &   $46.8\pm1.3$ &  7.1 & dwarf \\
20  & 18:24:24.1 & $-$24:43:36 &  8.64 & 1.16 &   $18.2\pm1.0$ &  8.8 & dwarf \\
21  & 18:24:57.4 & $-$24:44:21 & 11.15 & 0.85 &   $12.7\pm0.7$ &  9.7 & 7.296 \\
22  & 18:24:36.7 & $-$24:50:42 & 11.05 & 0.87 &   $14.6\pm0.7$ &  1.8 & 7.110 \\
23  & 18:24:38.7 & $-$24:50:21 & 11.69 & 0.82 &    $4.5\pm0.6$ &  2.3 & 6.678 \\
24  & 18:24:51.5 & $-$24:44:03 & 11.29 & 0.84 &    $2.5\pm0.8$ &  9.2 & 8.587 \\
25  & 18:24:41.1 & $-$24:48:25 & 11.03 & 0.82 &   $35.5\pm0.8$ &  4.2 & dwarf \\
26  & 18:24:37.7 & $-$24:48:52 & 11.92 & 0.81 &   $11.1\pm0.8$ &  3.5 & 6.683 \\
27  & 18:24:32.9 & $-$24:48:43 & 11.79 & 0.80 &   $32.6\pm1.0$ &  3.5 & dwarf \\
28  & 18:24:42.2 & $-$24:45:25 & 11.53 & 0.77 &   $39.8\pm0.8$ &  7.1 & dwarf \\
29  & 18:24:57.9 & $-$24:50:19 & 10.72 & 0.86 &  $-75.8\pm0.7$ &  6.3 & dwarf \\
31  & 18:24:46.7 & $-$24:48:53 & 10.46 & 0.92 &    $8.9\pm0.6$ &  4.7 & 7.583 \\
32  & 18:24:36.7 & $-$24:50:15 & 11.20 & 0.83 &  $-30.9\pm0.9$ &  2.2 & 8.739 \\
33  & 18:24:39.0 & $-$24:51:08 & 10.25 & 0.90 &   $20.4\pm0.8$ &  1.8 & 7.532 \\
34  & 18:24:35.0 & $-$24:52:04 &  8.92 & 1.05 &   $19.6\pm0.7$ &  0.5 & 8.571 \\
35  & 18:24:41.8 & $-$24:49:27 & 11.89 & 0.81 &    $2.5\pm0.7$ &  3.5 & 6.879 \\
36  & 18:24:37.4 & $-$24:49:07 & 11.32 & 0.86 &  $-50.2\pm1.3$ &  3.3 & dwarf \\
37  & 18:24:34.9 & $-$24:51:45 &  9.56 & 0.98 &   $12.0\pm0.6$ &  0.7 & 8.098 \\
38  & 18:24:37.9 & $-$24:50:51 & 10.52 & 1.05 &   $15.2\pm0.6$ &  1.8 & 7.563 \\
39  & 18:24:36.9 & $-$24:52:02 &  9.08 & 0.83 &   $19.0\pm0.7$ &  1.0 & 8.389 \\
40  & 18:24:56.8 & $-$24:48:14 & 11.17 & 0.83 &  $104.9\pm1.0$ &  6.9 & dwarf \\
41  & 18:24:42.0 & $-$24:52:20 & 11.55 & 0.82 &  $-35.0\pm0.7$ &  2.2 & 7.946 \\
42  & 18:25:06.4 & $-$24:52:04 & 10.32 & 0.90 &  $-22.2\pm1.2$ &  8.1 & dwarf \\
43  & 18:24:47.6 & $-$24:50:42 &  9.80 & 0.97 &   $15.7\pm0.7$ &  3.8 & 7.879 \\
44  & 18:25:12.1 & $-$24:50:28 & 11.16 & 0.83 &   $48.0\pm1.1$ &  9.6 & dwarf \\
45  & 18:24:49.4 & $-$24:52:08 & 11.60 & 0.83 &   $31.4\pm1.4$ &  4.0 & dwarf \\
47  & 18:24:42.0 & $-$24:51:30 & 10.20 & 0.91 &    $6.4\pm0.9$ &  2.3 & 7.751 \\
48  & 18:24:43.3 & $-$24:51:50 & 10.77 & 0.87 &   $16.8\pm0.7$ &  2.6 & 7.467 \\
49  & 18:24:56.3 & $-$24:50:09 & 10.39 & 0.91 &   $20.1\pm1.6$ &  6.0 & dwarf \\
50  & 18:24:51.9 & $-$24:54:43 & 11.41 & 0.80 & $-148.0\pm0.9$ &  5.3 & 6.886 \\
52  & 18:24:50.7 & $-$24:52:45 & 10.97 & 0.86 &   $12.6\pm1.3$ &  4.4 & dwarf \\
53  & 18:24:54.0 & $-$24:54:13 & 11.05 & 0.82 &  $-49.8\pm0.9$ &  5.5 & 8.681 \\
54  & 18:24:47.3 & $-$24:53:19 &  9.94 & 0.97 &   $14.9\pm0.6$ &  3.7 & 7.867 \\
55  & 18:24:35.9 & $-$24:52:09 & 10.91 & 0.86 &    $8.6\pm0.8$ &  0.7 & 7.330 \\
57  & 18:25:21.2 & $-$24:55:03 & 10.46 & 0.90 &  $-83.2\pm1.1$ & 12.2 & dwarf \\
58  & 18:25:04.1 & $-$24:53:25 & 11.68 & 0.78 &  $-61.3\pm1.1$ &  7.7 & dwarf \\
59  & 18:24:41.3 & $-$24:54:18 & 11.46 & 0.82 &   $10.4\pm0.7$ &  3.0 & 6.882 \\
60  & 18:24:41.5 & $-$24:54:34 & 11.13 & 0.86 &   $14.4\pm0.7$ &  3.2 & 7.132 \\
61  & 18:24:53.7 & $-$24:56:12 & 11.23 & 0.87 &    $4.8\pm0.7$ &  6.5 & 7.281 \\
62  & 18:24:36.9 & $-$24:52:42 & 10.89 & 0.87 &  $-4.7\pm0.7$ &  1.1 & 7.143 \\
63  & 18:24:43.3 & $-$24:56:17 &  8.69 & 1.14 &   $11.2\pm0.6$ &  4.8 & 8.720 \\
64  & 18:24:39.2 & $-$24:52:40 & 10.09 & 0.92 &   $15.6\pm0.8$ &  1.6 & 7.697 \\
65  & 18:24:49.8 & $-$24:56:29 & 10.37 & 0.91 &    $5.8\pm0.7$ &  6.0 & 7.633 \\
66  & 18:24:33.4 & $-$24:52:40 & 11.71 & 0.80 &   $10.6\pm0.8$ &  0.5 & 6.484 \\
67  & 18:24:51.8 & $-$24:54:00 &  8.70 & 1.11 &    $2.8\pm1.0$ &  5.0 & dwarf \\
68  & 18:24:52.7 & $-$24:55:12 & 11.54 & 0.80 &   $33.2\pm0.8$ &  5.7 & dwarf \\
69  & 18:24:37.7 & $-$24:53:08 & 10.89 & 0.87 &    $0.0\pm0.8$ &  1.5 & 7.209 \\
70  & 18:24:47.7 & $-$24:58:12 & 10.89 & 0.85 & $-114.5\pm1.1$ &  7.0 & dwarf \\
71  & 18:24:33.4 & $-$24:55:48 &  8.81 & 1.13 &    $6.2\pm0.6$ &  3.6 & 8.614 \\
72  & 18:24:36.3 & $-$24:57:24 & 11.69 & 0.81 &   $10.9\pm0.8$ &  5.3 & 6.734 \\
73  & 18:24:39.0 & $-$24:59:02 & 11.60 & 0.80 &  $-17.9\pm0.9$ &  7.0 & dwarf \\
74  & 18:24:36.2 & $-$24:54:47 & 10.68 & 0.90 &   $-0.5\pm0.7$ &  2.7 & 7.380 \\
75  & 18:24:33.9 & $-$24:54:07 & 11.84 & 0.78 &    $4.8\pm0.7$ &  2.0 & 6.727 \\
76  & 18:24:33.6 & $-$24:54:40 & 10.86 & 0.89 &   $11.9\pm0.7$ &  2.5 & 7.341 \\
77  & 18:24:45.6 & $-$24:58:44 &  9.80 & 0.98 &   $51.3\pm0.8$ &  7.3 & dwarf \\
78  & 18:24:39.4 & $-$24:54:52 & 11.24 & 0.82 &   $54.7\pm0.9$ &  3.1 & dwarf \\
79  & 18:24:34.4 & $-$24:53:16 & 11.96 & 0.78 &    $9.3\pm0.7$ &  1.1 & 6.466 \\
80  & 18:24:32.8 & $-$25:00:03 &  9.42 & 1.05 &    $8.7\pm0.7$ &  7.9 & 8.159 \\
81  & 18:24:21.0 & $-$25:04:17 & 10.86 & 0.87 &   $19.5\pm1.3$ & 12.4 & dwarf \\
82  & 18:24:27.7 & $-$24:58:42 & 11.31 & 0.85 &   $85.8\pm1.1$ &  6.6 & dwarf \\
83  & 18:24:31.3 & $-$24:52:47 &  9.76 & 0.97 &   $15.9\pm0.6$ &  0.7 & 8.024 \\
84  & 18:24:34.5 & $-$24:55:38 &  8.92 & 1.10 &   $11.0\pm0.7$ &  3.5 & 8.512 \\
85  & 18:24:26.4 & $-$24:56:31 & 11.51 & 0.83 &   $26.9\pm0.8$ &  4.6 & dwarf \\
86  & 18:24:27.7 & $-$24:56:26 & 10.67 & 0.83 &    $8.8\pm0.6$ &  4.4 & 6.890 \\
87  & 18:24:33.1 & $-$24:58:10 & 11.18 & 0.86 &   $54.6\pm1.2$ &  6.0 & dwarf \\
88  & 18:24:32.6 & $-$24:54:16 & 10.35 & 0.92 &    $5.5\pm0.7$ &  2.1 & 7.470 \\
89  & 18:24:05.8 & $-$24:56:40 &  9.02 & 1.07 &  $-12.0\pm0.7$ &  8.0 & 8.972 \\
90  & 18:23:57.8 & $-$24:58:49 & 11.32 & 0.82 &   $-8.5\pm1.2$ & 10.9 & 8.994 \\
91  & 18:24:30.8 & $-$24:52:36 &  9.40 & 1.04 &   $-8.7\pm0.6$ &  0.6 & 8.145 \\
92  & 18:24:12.6 & $-$24:56:58 & 10.87 & 0.85 &   $65.4\pm1.1$ &  6.9 & dwarf \\
93  & 18:24:25.9 & $-$24:53:57 & 11.32 & 0.83 &    $5.0\pm0.8$ &  2.4 & 6.898 \\
94  & 18:23:59.7 & $-$25:01:30 & 11.68 & 0.80 &  $-33.1\pm0.8$ & 12.4 & 8.181 \\
95  & 18:24:27.6 & $-$24:54:01 &  9.51 & 0.98 &    $8.5\pm0.6$ &  2.2 & 8.109 \\
96  & 18:24:26.5 & $-$24:51:47 & 10.92 & 0.85 &   $13.0\pm0.7$ &  1.6 & 7.282 \\
97  & 18:24:23.5 & $-$24:57:15 &  9.92 & 0.99 &  $-36.8\pm0.9$ &  5.6 & dwarf \\
98  & 18:24:07.8 & $-$25:03:00 & 10.95 & 0.86 &  $277.1\pm2.1$ & 12.5 & 6.335 \\
99  & 18:24:18.1 & $-$24:59:25 & 11.33 & 0.85 &   $89.9\pm0.9$ &  8.1 & dwarf \\
100 & 18:24:22.3 & $-$24:58:23 & 11.23 & 0.82 &   $67.8\pm0.9$ &  6.7 & dwarf \\
101 & 18:23:44.3 & $-$24:54:43 & 10.89 & 0.87 &   $27.3\pm1.4$ & 12.1 & dwarf \\
102 & 18:23:51.8 & $-$24:54:31 & 11.92 & 0.80 &  $-11.8\pm0.8$ & 10.3 & 8.019 \\
103 & 18:24:23.7 & $-$24:52:34 & 10.71 & 0.86 &    $8.5\pm0.7$ &  2.2 & 7.403 \\
104 & 18:24:05.5 & $-$24:55:11 &  9.19 & 1.05 &   $13.3\pm0.7$ &  7.3 & 8.341 \\
105 & 18:24:29.9 & $-$24:52:31 & 11.93 & 0.79 &    $9.2\pm0.7$ &  0.8 & 6.674 \\
106 & 18:24:29.9 & $-$24:52:18 & 11.41 & 0.79 &    $5.0\pm0.7$ &  0.7 & 6.972 \\
107 & 18:23:48.7 & $-$24:58:22 & 10.52 & 0.90 &  $-54.7\pm1.2$ & 12.5 & dwarf \\
108 & 18:24:13.0 & $-$24:55:02 & 10.81 & 0.85 &  $-17.3\pm1.2$ &  5.6 & dwarf \\
109 & 18:24:01.8 & $-$24:50:48 & 11.77 & 0.77 &  $128.2\pm1.1$ &  7.6 & dwarf \\
110 & 18:24:10.4 & $-$24:51:21 & 11.32 & 0.87 &  $-10.2\pm1.0$ &  5.5 & 8.302 \\
111 & 18:23:57.5 & $-$24:52:09 & 10.87 & 0.86 &   $-5.0\pm1.0$ &  8.5 & dwarf \\
112 & 18:23:42.3 & $-$24:52:00 & 10.74 & 0.90 &   $36.5\pm1.2$ & 12.2 & dwarf \\
113 & 18:24:00.0 & $-$24:52:26 & 11.96 & 0.79 &  $-58.5\pm0.9$ &  7.9 & dwarf \\
114 & 18:23:48.6 & $-$24:52:31 & 11.20 & 0.86 &    $8.9\pm0.9$ & 10.7 & dwarf \\
115 & 18:24:23.5 & $-$24:51:56 & 10.06 & 0.96 &   $-4.3\pm0.7$ &  2.3 & 7.806 \\
116 & 18:24:02.2 & $-$24:53:36 &  9.76 & 0.93 &   $62.0\pm0.7$ &  7.6 & dwarf \\
117 & 18:23:43.8 & $-$24:54:21 &  9.51 & 1.00 &   $12.5\pm0.9$ & 12.1 & dwarf \\
118 & 18:24:11.2 & $-$24:47:31 & 11.67 & 0.80 &   $36.0\pm0.9$ &  6.9 & dwarf \\
119 & 18:24:04.7 & $-$24:47:24 & 11.36 & 0.81 &   $24.3\pm0.8$ &  8.2 & dwarf \\
123 & 18:24:13.1 & $-$24:50:04 & 11.11 & 0.85 &   $53.3\pm1.4$ &  5.2 & dwarf \\
125 & 18:23:58.0 & $-$24:48:13 & 10.74 & 0.85 &    $6.0\pm0.6$ &  8.8 & 7.371 \\
\hline
\end{longtable}

\end{document}